\documentclass[conference]{IEEEtran}
\IEEEoverridecommandlockouts

\usepackage[utf8]{inputenc}
\usepackage{amsmath,amssymb,amsthm}
\usepackage{enumerate}
\usepackage{stfloats}
\usepackage{comment}
\usepackage{subcaption}
\usepackage{siunitx}
\usepackage{mathtools}
\usepackage{accents,latexsym,cancel}
\usepackage{cite}

\usepackage[dvipsnames]{xcolor}
\definecolor{myPink}{RGB}{255,105,183}

\usepackage[T1]{fontenc}
\usepackage{graphics} % for pdf, bitmapped graphics files
\usepackage{epsfig} % for postscript graphics files
\usepackage[mathscr]{euscript}
\usepackage{algorithm}
\usepackage[noend]{algpseudocode}
\usepackage{bbm}
\makeatletter
\def\BState{\State\hskip-\ALG@thistlm}
\makeatother

\usepackage{tikz}
\usetikzlibrary{arrows,shapes,chains,matrix,positioning,scopes,patterns,calc,
decorations.markings,
decorations.pathmorphing,
}

\usepackage{pgfplots}
\pgfplotsset{compat=1.3}
\usepgflibrary{shapes}

\newtheorem{theorem}{Theorem}
\newtheorem{lemma}[theorem]{Lemma}
\newtheorem{condition}[theorem]{Condition}
\newtheorem{proposition}[theorem]{Proposition}
\newtheorem{definition}[theorem]{Definition}

\newtheorem{remark}[theorem]{Remark}
\newtheorem{corollary}[theorem]{Corollary}

\renewcommand{\epsilon}{\varepsilon}

\newcommand{\RNum}[1]{\uppercase\expandafter{\romannumeral #1\relax}}

\newcommand{\bv}{\ensuremath{\mathbf{b}}}

\newcommand{\ev}{\ensuremath{\mathbf{e}}}

\newcommand{\rv}{\ensuremath{\mathbf{r}}}
\newcommand{\Rv}{\ensuremath{\mathbf{R}}}
\newcommand{\sv}{\ensuremath{\mathbf{s}}}
\newcommand{\Sv}{\ensuremath{\mathbf{S}}}

\newcommand{\vv}{\ensuremath{\mathbf{v}}}

\newcommand{\wv}{\ensuremath{\mathbf{w}}}
\newcommand{\xv}{\ensuremath{\mathbf{x}}}

\newcommand{\yv}{\ensuremath{\mathbf{y}}}

\newcommand{\zv}{\ensuremath{\mathbf{z}}}

\newcommand{\etav}{\ensuremath{\boldsymbol{\eta}}}

\newcommand{\zerov}{\ensuremath{\boldsymbol{0}}}

\newcommand{\Am}{\ensuremath{\mathbf{A}}}

\newcommand{\IDm}{\ensuremath{\mathbf{I}}}

\def\Pr{\mathrm{Pr}}

\DeclareMathAlphabet{\mcl}{OMS}{cmsy}{m}{n}

\newlength\tikzwidth
\newlength\tikzheight

\definecolor{mycolor1}{rgb}{0.63529,0.07843,0.18431}%
\definecolor{mycolor2}{rgb}{0.00000,0.44706,0.74118}%
\definecolor{mycolor3}{rgb}{0.00000,0.49804,0.00000}%
\definecolor{mycolor4}{rgb}{0.87059,0.49020,0.00000}%
\definecolor{mycolor5}{rgb}{0.00000,0.44700,0.74100}%
\definecolor{mycolor6}{rgb}{0.74902,0.00000,0.74902}%

\title{On Sparse Regression LDPC Codes}
\author{Jamison R. Ebert, Jean-Francois Chamberland, Krishna R. Narayanan\\
Department of Electrical and Computer Engineering, Texas A\&M University
\thanks{
This material is based upon work supported, in part, by the National Science Foundation (NSF) under Grant CCF-2131106, and by Qualcomm Technologies, Inc., through their University Relations Program.}
}

\begin{document}

\maketitle

\begin{abstract}

Belief propagation applied to iterative decoding and sparse recovery through approximate message passing (AMP) are two research areas that have seen monumental progress in recent decades.
Inspired by these advances, this article introduces sparse regression LDPC codes and their decoding.
Sparse regression codes (SPARCs) are a class of error correcting codes that build on ideas from compressed sensing and can be decoded using AMP.
In certain settings, SPARCs are known to achieve capacity; yet, their performance suffers at finite block lengths.
Likewise, LDPC codes can be decoded efficiently using belief propagation and can also be capacity achieving.
This article introduces a novel concatenated coding structure that combines an LDPC outer code with a SPARC-inspired inner code.
% A typical LDPC code is generated over a finite field, most notably the binary field; as such, LDPC-coded symbols are often mapped to a signal constellation in a process that can be tedious and inherently constrains performance.
% To circumvent these issues, this article introduces a novel concatenated coding structure that combines an LDPC outer code with a SPARC-inspired inner code.
Efficient decoding for such a code can be achieved using AMP with a denoiser that performs belief propagation on the factor graph of the outer LDPC code.
% This concatenated structure conforms to the theory of state evolution for non-separable denoisers and therefore is amenable to analysis.
The proposed framework exhibits performance improvements over SPARCs and standard LDPC codes for finite block lengths and results in a steep waterfall in error performance, a phenomenon not observed in uncoded SPARCs.
% Some sparse regression LDPC codes are surprisingly amenable to analysis, with performance guarantees and suitable criteria for code optimization.
Findings are supported by numerical results.

\end{abstract}


\section{Introduction and Background}
\label{section:Introduction}

Low-density parity check (LDPC) codes have been the subject of many scientific inquiries in the past, with landmark contributions ranging from theoretical advances to practical implementations~\cite{gallager1962low,mackay1999good,luby2001improved,richardson2001capacity,chung2001analysis,davey1998low,bennatan2006design,richardson2008modern}.
LDPC ensembles are amenable to analysis and, under certain conditions, encoded messages can be recovered efficiently using iterative belief propagation (BP) decoding.
Interestingly, the complexity of BP decoding grows linearly with block length and, as such, this paradigm offers a pragmatic solution for long block lengths~\cite{costello2014spatially}.
Some spatially coupled LDPC constructions feature capacity approaching iterative decoding thresholds while also avoiding the pitfall of error floors~\cite{felstrom1999time,lentmaier2010iterative,kudekar2013spatially,yedla2014simple,kumar2014threshold,andriyanova2016threshold}.
Unfortunately, systems operating at shorter block lengths may not be conducive to the application of spatial coupling.
In such situations, non-binary LDPC codes have been leveraged as a means to provide adequate performance~\cite{davey1998low,bennatan2006design,declercq2007decoding,voicila2010low,chang2012non}.
% It is also worth mentioning that the advent of polar codes has shifted the coding landscape for practical systems~\cite{arikan2009channel}.

In a seemingly unrelated research direction, Joseph and Barron introduced the concept of a sparse regression code (SPARC), which establishes a connection between code design and sparse recovery in high dimensions~\cite{joseph2012least,joseph2013fast,venkataramanan2019sparse}.
The codewords contained in a SPARC codebook consist of sparse superpositions of columns from a design matrix, also known as a sensing matrix in the compressed sensing (CS) literature~\cite{donoho2006compressed,candes2006robust}.
In some sense, the SPARC decoding process is equivalent to recovering the non-zero entries in a suitably designed index vector.
Low complexity algorithms have been proposed for the efficient decoding of SPARC codewords.
Notably, Rush \textit{et al.}\ discuss an approximate message passing (AMP) decoder for SPARCs in~\cite{rush2017capacity,rush2021capacity}.
It is also worth mentioning that AMP has independently gained much research momentum in recent years owing to its low implementation cost, mathematical tractability, and excellent performance~\cite{donoho2009message,bayati2011dynamics,javanmard2013state,bayati2015universality,berthier2020state}.
Sparse regression codes paired with AMP decoding are known to achieve asymptotic single-user AWGN channel capacity under an appropriately chosen power allocation.
Concurrently, there have been efforts to improve the finite block-length performance of SPARCs~\cite{greig2017techniques,cao2021using}.

A popular strategy to improve the performance of codes in practical settings is to adopt a concatenated structure.
For example, SPARCs and LDPC codes have been paired in the past to create concatenated schemes that improve performance~\cite{greig2017techniques,liang2020compressed,liu2021capacity,cao2021using}.
Concatenated structures with a SPARC-like inner code have also been proposed in unsourced random access~\cite{fengler2021sparcs,amalladinne2022unsourced,ebert2022coded}.
In~\cite{amalladinne2022unsourced}, it is shown that the structure of a judiciously designed outer code can be integrated into the denoising function of the inner AMP recovery algorithm.
Surprisingly, despite being mentioned by Liu \textit{et al.}\ in~\cite{liu2021capacity} as a possible future research direction, such an approach has not been considered for the single-user scenario.
This article seeks to address this defficiency by introducing sparse regression LDPC codes. 
% We seeks to address this deficiency by considering the benefits of a dynamic denoiser in the context of sparse regression LDPC codes.
% There appears to be a natural connection between a SPARC inner code and a non-binary LDPC outer code where the field size of the outer code matches the section length of the SPARC inner code.
% In such cases, BP on the factor graph of the outer LDPC code arises as an natural denoising function for AMP decoding, striking a balance between performance and complexity.
% As predicted in~\cite{liu2021capacity}, state evolution for the ensuing AMP algorithm relies on recent results for non-separable functions~\cite{berthier2020state}.
% 
% In this context, the goal of this article is to devise a coding architecture wherein the structure of the outer code informs the design of an AMP algorithm for inner decoding.

Sparse regression LDPC codes feature a concatenated inner SPARC-like code and an outer non-binary LDPC code. 
This concatenated structure is amenable to a decoder that exchanges information between inner and outer codes via a dynamic AMP denoiser. 
In this setting, the denoiser performs BP on the factor graph of the outer LDPC code to improve the AMP state estimates.
This in turn improves the effective observations which in turn translate into better local belief vectors for the BP LDPC decoder. 
Such a denoiser improves AMP's convergence and lowers the residual mean square error (MSE) in the estimated SPARC codeword.
Once the MSE of the iterative AMP-BP decoding levels off, the outer LDPC code can be leveraged once more to further improve performance. 
% Once the MSE of the AMP-BP decoding levels off, the effective observations can be processed by a standard LDPC decoder as a last step.
The goal is to use AMP-BP decoding to increase the SNR of the effective observation to a level that exceeds the BP threshold of the graph and then to apply message passing to the outer LDPC code to recover the message.
Numerical results demonstrate the performance advantages of the proposed model over related architectures at reasonably short block lengths.
% Concurrently, the envisioned approached may offer an efficient way to implement constellation shaping, with channel input being Gaussian-like, and it may improve performance for short block-lengths applications.

%Iterative decoding thresholds get further from capacity as the graph density increases.

\section{System Model} 
\label{section:SystemModel}

% The mathematical setting for the problem at hand is a familiar one.
We consider a point-to-point memoryless additive white Gaussian noise (AWGN) channel with one transmitter and one receiver, each equipped with a single antenna.
The received signal $\yv \in \mathbb{R}^n$ is given by
\begin{equation} \label{equation:ChannelModel}
\yv = \xv + \zv,
\end{equation}
where $\xv \in \mathbb{R}^n$ denotes the transmitted signal and $\zv \sim \mathcal{N}\left(0, \sigma^2\IDm\right)$ represents AWGN. 
% The transmitted signal must satisfy an average power constraint $\mathbb{E}\left[\|\xv\|_2^2\right] \leq nP$, where $n$ denotes the number of channel uses. 
% The noise is a random vector composed of a series of independent Gaussian random variables, each with mean zero and variance $\sigma^2$.
% The set of candidate codewords is subject to an average power constraint.
% Parameter $n$ denotes the number of channel uses or, equivalently, the number of (real) degrees of freedom available.
% The signal-to-noise ratio (SNR) is expressed as
% \begin{equation} \label{equation:SNR}
% \operatorname{SNR} = \frac{\mathbb{E} \left[ \| \Xv \|^2 \right]}{2 B \sigma^2} .
% \end{equation},
% where $B$ denotes the number of bits. 
% Without loss of generality, the power constraint can be set to one, i.e. $\mathbb{E}\left[\|\Xv\|^2\right] = 1$, with the understanding that a distinct power level can be absorbed in the effective noise variance.
As mentioned above, we wish to study a coding architecture composed of a sparse regression inner code~\cite{joseph2012least,joseph2013fast,venkataramanan2019sparse}, and a non-binary LDPC outer code~\cite{davey1998low,bennatan2006design,richardson2008modern}.
We elaborate on the encoding process and the decoding scheme below.

\subsection{Sparse Regression LDPC Encoding}

The proposed encoding process features a sequence of three distinct steps: $q$-ary LDPC encoding, indexing LDPC symbols, and inner CS encoding.
First, the information bits are encoded into a $q$-ary LDPC codeword via well-established operations~\cite{davey1998low,bennatan2006design,richardson2008modern}.
The second step transforms the $q$-ary LDPC codeword into a suitable sparse vector.
The last step is the matrix multiplication emblematic of a sparse regression code; the output is sometimes referred to as a large random matrix system~\cite{joseph2012least,joseph2013fast,venkataramanan2019sparse,liu2021capacity}.
We summarize the notions pertaining to this encoding process below while concurrently introducing necessary notation.

\subsubsection{$q$-ary LDPC Encoding}

% The envisioned architecture warrants the adoption of a non-binary LDPC outer code.
The LDPC encoder takes a binary sequence $\wv \in \mathbb{F}_2^{B}$ as its input and maps it to a $q$-ary codeword $\vv \in \mathbb{F}_q^L$.
We represent the resultant codeword in concatenated form as
\begin{equation} \label{equation:CodewordLDPC2}
\vv = \left( v_1, v_2, \ldots, v_L \right)
\end{equation}
where the $\ell$th element $v_{\ell}$ lies in finite field $\mathbb{F}_{q}$ and $L$ is the length of the codeword.

% For our upcoming discussion, it is useful to impose a total order on the elements of $\mathbb{F}_q$.
% Within this finite order, the zero element of the field appears first and the one element of the field comes next.
% The position for the remaining elements is somewhat arbitrary, but fixed.
% For example, a natural order for $\mathbb{F}_2^v$ is to map a binary string to its integer representation when interpreted as an integer in binary form (positional numeral system with radix two).
% Returning to the general case, the chosen order induces a bijection between the field elements in $\mathbb{F}_q$ and the sequence of integers
% \begin{equation*}
% [q] = 0, 1, \ldots, q-1 .
% \end{equation*}

Note that there exists a bijection $\Phi: \mathbb{F}_q \rightarrow [q]$ between the elements of $\mathbb{F}_q$ and the integers $[q] = ( 0, 1, \ldots, q-1 )$, where $0 \in \mathbb{Z}$ maps to $0 \in \mathbb{F}_q$ and $1 \in \mathbb{Z}$ maps to $1 \in \mathbb{F}_q$.
Throughout, we adopt such an arbitrary, but fixed bijection and exploit this relation by employing the same variable for a field element $g \in \mathbb{F}_q$ and for its corresponding integer $\Phi\left(g\right) \in [q]$. 
This slight abuse of notation greatly simplifies exposition, and its use should not lead to confusion because one can unambiguously infer from context whether an instance refers to the field element or to its integer representation. 
% Another conceptual justification for our approach is to think of functions that act on both integers and field elements as overloaded.
% When faced with a field element as an argument, such a function first calls the bijection $\Phi$ to get a suitable integer and, then, returns the appropriate output.

\subsubsection{LDPC Symbol Indexing}

% With Remark~\ref{remark:NotationOverload} in mind, it becomes straightforward to explain the second step of the encoding process.
% Recall that each $v_{\ell} \in \vv$ is an element of finite field $\mathbb{F}_q$.
Coded symbol sparsification/indexing consists of mapping $v_{\ell} \in \mathbb{F}_q$ to standard basis vector $\ev_{v_{\ell}} \in \mathbb{R}^{q}$ and subsequently stacking the $L$ basis vectors together.
We seize this opportunity to reinforce the notion that entry $\ell$ of $\vv$ is an element of $\mathbb{F}_q$, but $v_{\ell}$ in $\ev_{v_{\ell}}$ refers to an integer in $[q]$ under our overloaded notation. 
With that, the output of the indexing process becomes
\begin{equation} \label{equation:Indexing}
\sv = \begin{bmatrix} \ev_{v_1} \\ \vdots \\ \ev_{v_L} \end{bmatrix} ,
\end{equation}
where $\sv$ is an $L$-sparse vector of length $qL$.
Vector $\sv$ has a structure akin to that of a sparse regression code prior to multiplication by a large random matrix.
This structured sparsity can be exploited during decoding.

\subsubsection{Inner CS Encoding}

The last phase of the encoding process consists in pre-multiplying vector $\sv$ by matrix $\Am$ to get $\xv = \Am \sv$, where $\Am \in \mathbb{R}^{n \times q L}$ and $\Am_{i, j} \sim \mathcal{N}\left(0, \frac{1}{n}\right)$.
Note that $n \ll q L$ to conform to the CS framework. 
% Conforming to the parameters of our system, $\Am$ has dimensions $n \times q L$, where $n \ll q L$.
% The entries of $\Am$ are generated randomly and independently, each with Gaussian distribution $\mathcal{N}(0, 1/n)$.
Equation \eqref{equation:ChannelModel} may thus be rewritten as:
\begin{equation} \label{equation:UpdatedChannelModel}
\yv = \Am \sv + \zv .
\end{equation}
Having specified the code structure, we are now ready to discuss the process of recovering $\sv$ from $\yv$.
% The overall encoding process is depicted in Fig.~\ref{figure:EncodingProcess}.
% \begin{figure}
% \centering
% \input{Figures/EncodingProcess}
% \caption{This diagram depicts the encoding process for an SRLDPC code.
%     Information message $\wv$ is first outer encoded using an LDPC code over $\mathbb{F}_q$. 
%     Every LDPC-encoded field element is subsequently converted into a one-sparse basis vector.
%     The collection of one-sparse vectors are then stacked into a SPARC-like sequence.
%     The resulting $L$-sparse vector is pre-multiplied by matrix $\Am$, yielding signal $\xv$.}
% \label{figure:EncodingProcess}
% \end{figure}

\subsection{Sparse Regression LDPC Decoding}

Paralleling the development of AMP for sparse regression codes~\cite{rush2017capacity} and drawing inspiration from concatenated AMP systems~\cite{liang2020compressed,liu2021capacity}, we wish to create a composite iterative process to recover state vector $\sv$ from $\yv$ that exploits the sparsity in $\sv$ and the parity structure of the LDPC code.
% However, a notable distinction between our system model and previously published articles is the presence of a $q$-ary LDPC outer code.
% Thus, we wish to create an AMP decoder that simultaneously takes advantage of the sparsity in $\sv$ and the parity structure embedded in the LDPC outer code.
This can be accomplished by incorporating message passing on the factor graph of the LDPC code into the AMP denoiser.
A similar approach can be found in~\cite{amalladinne2022unsourced}; however, the denoiser we wish to utilize below differs in the fact that only one codeword is present within $\yv$.
% This distinction simplifies the structure of the code and enables us to leverage a denoiser that more closely parallels traditional message passing algorithms for $q$-ary LDPC codes.

Our AMP composite algorithm is as follows,
\begin{align}
\rv^{(t)} &= \Am^{\mathrm{T}} \zv^{(t)} + \sv^{(t)} \label{equation:EffectiveObservation} \\
\zv^{(t)} &= \yv - \Am \sv^{(t)} + \frac{\zv^{(t-1)}}{n} \operatorname{div} \etav_{t-1} \left( \rv^{(t-1)} \right) \label{equation:AMP-Residual} \\
\sv^{(t+1)} &= \etav_{t} \left( \rv^{(t)} \right) \label{equation:AMP-Denoising}
\end{align}
where the superscript $t$ denotes the iteration count.
The algorithm is initialized with conditions $\rv^{(0)} = \sv^{(0)} = \zerov$ and $\zv^{(0)} = \yv$.
% By convention, every quantity with a negative iteration count is equal to the zero vector.

Equation~\eqref{equation:EffectiveObservation} specifies the \emph{effective observation}, which is characteristic of AMP.
Equation \eqref{equation:AMP-Residual} computes the \textit{residual} error enhanced with an Onsager correction term,
whereas \eqref{equation:AMP-Denoising} provides a \emph{state estimate}.
The denoising functions $\left( \etav_t (\cdot) \right)_{t \geq 0}$ seek to exploit the structure of $\sv$ while computing the state updates.
% \begin{figure*}[t]
% \centering
% \input{Figures/DecoderAMP2}
% \caption{This notional diagram depicts the operation of the dynamic AMP-BP decoder.
% The input comes in the form of observation $\yv$ at the top left.
% During every AMP iteration, the algorithm computes the residual $\zv$, which incorporates the effect of the Onsager term.
% This vector is then turned into an effective observation, which acts as the input to the BP denoiser.
% After message passing, a state estimate vector is produced for every section.
% This, in turn, produces the updated global estimate via concatenation.
% The computation of the Onsager contribution, which is intrinsic to AMP, is highlighted on the left.
% The iterative process repeats itself until convergence is achieved, at which point the state estimate $\hat{\sv}$ is taken as the output of the algorithm.
% }
% \label{figure:DecodingProcess}
% \end{figure*}
% The argument of the denoiser in \eqref{equation:AMP-Denoising} is called the \emph{effective observation}.

Generally speaking, a choice denoiser is the conditional expectation of $\sv$ given $\rv^{(t)}$,
Unfortunately, this approach is computationally intractable in the present context because it entails summing over all the possible codewords.
As an alternative, we know that BP can be efficiently applied to $q$-ary LDPC codes. 
Furthermore, at any point during BP, a belief on individual LDPC symbols can be formed based on incoming messages from neighboring factor nodes and local observations. 
Thus, we can potentially apply a few iterations of BP as a means to get an estimate for the state vector by leveraging the connection between sections of $\sv$ and LDPC symbols.
In this sense, iterative message passing can act as a foundation for pragmatic denoising functions.
We elaborate on this connection below and, concurrently, we review pertinent notions of BP applied to $q$-ary LDPC codes.

\section{Creating a BP Deonoiser for AMP}
\label{section:BPDenoiser}

In this section, we introduce the denoising function we wish to employ within AMP.
% As a preliminary step, we examine the character of the effective observation $\rv$, given in \eqref{equation:EffectiveObservation}.
To begin, we emphasize that $\rv$ admits a sectionized representation akin to that of the state vector $\sv$ in \eqref{equation:Indexing}.
That is, we can view both the state estimate and the effective observation as concatenations of $L$ vectors, each of length $q$.
% Likewise, the state estimate $\hat{\sv}$ can be partitioned in a block fashion.
Mathematically, we have
\begin{xalignat*}{2}
\rv &= \begin{bmatrix} \rv_1 \\ \vdots \\ \rv_L \end{bmatrix} &
\hat{\sv} &= \begin{bmatrix} \hat{\sv}_1 \\ \vdots \\ \hat{\sv}_L \end{bmatrix} .
\end{xalignat*}
This point is crucially important because the denoiser is constructed in a block-wise fashion.
As a side note, we neglect the superscript $(t)$ for most of the discussion below to lighten notation.
% In general, the statements we make below apply to every iteration within the decoding process; as such, the compact notation should not lead to any confusion.
Furthermore, we emphasize that the presence of the hat in $\hat{\sv}$ distinguishes the estimate from the true state vector $\sv$ in the absence of an iteration count $(t)$.
% To help keep track of variables, Fig.~\ref{figure:DecodingProcess} illustrates several of the key quantities we use throughout.

Each section $\rv_{\ell}$ in $\rv$ acts as a vector observation about the value of $\sv_{\ell} \in \left\{ \ev_g : g \in [q] \right\}$.
An astounding and enabling property of AMP is that, under certain technical conditions, the effective observation $\rv$ is asymptotically distributed as $\sv + \tau \boldsymbol{\zeta}$, where $\boldsymbol{\zeta}$ is a random vector with independent $\mathcal{N}(0,1)$ components and $\tau$ is a deterministic quantity.
This fact hinges on the presence of the Onsager term in \eqref{equation:AMP-Residual} and on some smoothness conditions for the denoising functions. 
%The Onsager term is derived in Appendix \ref{appendix:OnsagerDerivation} and the required smoothness conditions are demonstrated in Appendix \ref{appendix:LipschitzDenoiser}. 
% These technical conditions are treated in Appendix \ref{appendix:LipschitzDenoiser}. 
% While we delay the treatment of these technical conditions to the appendices, we take advantage of the Gaussian distribution in our discussion below.
We delay the treatment of these technical conditions for the time being;
rather, we posit the requirements and formally introduce them as a condition.

\begin{condition} \label{condition:AsymptoticCharacterization}
The effective observation $\rv^{(t)}$ is asymptotically distributed as $\sv + \tau_t \boldsymbol{\zeta}_t$, where $\boldsymbol{\zeta}_t$ is a random vector with independent $\mathcal{N}(0,1)$ components and $\tau_t$ is specified by a set of deterministic equations.
This asymptotic characterization takes place in the dimensions of the system, as opposed to time or iteration count.
\end{condition}

We describe below our rationale behind the denoising function assuming Condition~\ref{condition:AsymptoticCharacterization} holds; we provide a rigorous foundation for this condition in Appendix \ref{appendix:LipschitzDenoiser}, but this can only be done once the structure of the denoiser is established.
Consider the effective observation restricted to section~$\ell$.
Under Condition~\ref{condition:AsymptoticCharacterization}, the distribution of random observation vector $\Rv_{\ell}$ given section $\Sv_{\ell} = \ev_g$ is given by
\begin{equation*}
f_{\Rv_{\ell} | \Sv_{\ell}} \left( \rv_{\ell} | \ev_g \right)
= \frac{1}{(2 \pi)^{\frac{q}{2}} \tau^q }
\exp \left( - \frac{\left\| \rv_{\ell} - \ev_g \right\|^2}{2 \tau^2} \right) .
\end{equation*}
% It may be beneficial to think of the inner AMP loop as being equivalent to using a Gaussian vector channel $L$ times, with every channel use being attached to one LDPC symbol.
Under a uniform input distribution, the conditional probability of $\Sv_{\ell} = \ev_g$ becomes
\begin{equation} \label{equation:ConditionalProbability}
\begin{split}
&\boldsymbol{\alpha}_{\ell} (g)
= \Pr \left( \Sv_{\ell} = \ev_g \middle| \Rv_{\ell} = \rv_{\ell} \right)
= \Pr \left( V_{\ell} = g \middle| \Rv_{\ell} = \rv_{\ell} \right) \\
&= \frac{f_{\Rv_{\ell} | V_{\ell}} \left( \rv_{\ell} | g \right)}
{\sum_{h \in \mathbb{F}_q} f_{\Rv_{\ell} | V_{\ell}} \left( \rv_{\ell} | h \right)}
= \frac{e^{- \frac{\left\| \rv_{\ell} - \ev_g \right\|^2}{2 \tau^2}}}
{\sum_{h \in \mathbb{F}_q} e^{- \frac{\left\| \rv_{\ell} - \ev_h \right\|^2}{2 \tau^2}}} \\
% &= \frac{e^{- \frac{\left( \rv_{\ell}(g) - \ev_g \right)^2 - \left( \rv_{\ell}(g) \right)^2}{2 \tau^2}}}
% {\sum_{h \in \mathbb{F}_q} e^{- \frac{\left( \rv_{\ell}(h) - \ev_h \right)^2 - \left( \rv_{\ell}(h) \right)^2}{2 \tau^2}}}
&= \frac{e^{\frac{\rv_{\ell}(g)}{\tau^2}}}
{\sum_{h \in \mathbb{F}_q} e^{\frac{\rv_{\ell}(h)}{\tau^2}}} .
\end{split}
\end{equation}
% Equivalently, one can form a likelihood for every possible field element 
% \begin{equation} \label{equation:LocalObservation}
% \begin{split}
% \mathcal{L}_{v_{\ell}} \left( g | \rv_{\ell} \right)
% &= f_{\Rv_{\ell} | \Sv_{\ell}} \left( \rv_{\ell} | \ev_g \right)
% = \frac{1}{(2 \pi)^{\frac{q}{2}} \tau^q }
% e^{- \frac{\left\| \rv_{\ell} - \ev_g \right\|^2}{2 \tau^2}} .
% \end{split}
% \end{equation}
% This system appears in Fig.~\ref{figure:FactorGraph}.
% The reader may notice that vector entries are repeatedly indexed by field elements; this usage conforms to the overloaded notation discussed in Remark~\ref{remark:NotationOverload}.
% \textcolor{orange}{Recall that we have imposed a finite order on $\mathbb{F}_q$, which induces a bijection between field elements in $\mathbb{F}_q$ and $[q]$.
% The value of $\rv_{\ell} (g)$ with $g \in \mathbb{F}_q$ is given by the $k$th entry in $\rv_{\ell}$, where $k$ is the position of field element $g$ in the aforementioned order.}

% \begin{figure}[t]
%   \centering
%   \input{Figures/FactorGraph}
%   \caption{
%   This illustration shows the augmented factor graph for the denoising function with the variable nodes, the parity check constraints, and the extra factors associated with local observations.
%   The effective observation vector is sectionized in a way that matches variable nodes.}
%   \label{figure:FactorGraph}
% \end{figure}

A possible estimate for $\Sv_{\ell}$ can be computed by taking its conditional expectation, given observation $\Rv_{\ell} = \rv_{\ell}$, with
\begin{equation} \label{equation:LocalMMSE}
\mathbb{E} \left[ \Sv_{\ell} \middle| \Rv_{\ell} = \rv_{\ell} \right]
= \sum_{g \in \mathbb{F}_q} \ev_g \Pr \left( \Sv_{\ell} = \ev_g \middle| \Rv_{\ell} = \rv_{\ell} \right) .
\end{equation}
A variant of this approach can be found in \cite{rush2017capacity} for a system without an outer code.
% It is also employed in \cite{fengler2019sparcs} in the context of unsourced random access.
% Yet, this approach overlooks the redundancy found in the outer code for the system under consideration.
Ideally, we would like to take advantage of the outer code with the more precise MMSE estimate of the form $\mathbb{E} \left[ \Sv_{\ell} \middle| \Rv = \rv \right]$.
Unfortunately, as mentioned above, computing this conditional expectation is far too complex to be implemented in practice.
A viable alternative that trades off performance and complexity is to execute BP on the factor graph of the outer LDPC code.
Implicitly, this approach computes an estimate for every $\Sv_{\ell}$ based on the computation tree of the code, up to a certain depth~\cite{wiberg1995codes,richardson2008modern}.
% We expound on the soft decoding of the LDPC outer code in the next section.

Frameworks to perform BP on factor graphs are well-established~\cite{kschischang2001factorgraph}; thus, we assume some familiarity with such iterative procedures. 
% Accordingly, we assume some familiarity with such iterative procedures.
% The main departures of our work from traditional iterative LDPC decoding comes from the non-binary finite field employed for the code and, more importantly, from the fact that we wish to integrate BP iterations within an AMP denoiser.
% Indeed, we want the AMP inner decoder and the BP outer decoder to operate in conjunction and promote convergence towards the desired solution, as opposed to applying these two decoders in a disjoint succession.
% 
% Unfortunately, notation for the treatment of BP decoding of $q$-ary LDPC codes and AMP is heavy.
% Throughout, we seek to adopt notations that mimic established practices in these two areas while avoiding overlaps.
% The same can be said for the analysis of AMP algorithms.
% hen the two are merged, the notation rapidly becomes overwhelming.
% To alleviate this challenge, we try to adopt notations that mimic established practices in these two distinct areas, while avoiding overlaps.
% Specifically, for the $q$-ary LDPC portion of the article, we borrow definitions and concepts from Bennatan and Burshtein~\cite{bennatan2006design}.
We proceed by first considering the nuances of non-binary LDPC factor graphs, then presenting a BP algorithm, and finally proposing a dynamic denoiser for SR-LDPC codes.

\subsection{Non-Binary LDPC Graphs}

The factor graph for an $\mathbb{F}_q$ LDPC code features $L$ variable (left) nodes and $L (1 - R)$ check (right) nodes enforcing parity constraints, where $R$ is the \emph{rate} of the LDPC code~\cite{davey1998low,bennatan2006design,richardson2008modern}.
An important distinction between binary and non-binary LDPC codes is that a factor graph for the latter code typically includes edge labels.
These labels take values in $\mathbb{F}_q \setminus \{ 0 \}$.
% Fig.~\ref{figure:FactorGraph} depicts a notional factor graph for a non-binary LDPC code, where the edge labels are represented as dots along the graph edges.
A vector $\vv \in \mathbb{F}_q^L$ is a valid codeword if it satisfies the parity equations
\begin{equation} \label{equation:ParityConstraint}
\sum_{v_{\ell} \in N(c_p)} \omega_{\ell, p} \otimes v_{\ell} = 0 \qquad \forall p \in [L (1 - R)],
\end{equation}
where $N(c_p)$ is the collection of variable nodes adjacent to parity check~$c_p$.
The summation and the multiplication operator $\otimes$ in \eqref{equation:ParityConstraint} take place over finite field $\mathbb{F}_q$.
Parameter $\omega_{\ell, p}$ represents the label or weight assigned with the edge connecting variable node $v_{\ell}$ and check node $c_p$.
Adopting common factor graph concepts~\cite{kschischang2001factorgraph,loeliger2004introduction}, we denote the graph neighbors of variable node $v_{\ell}$ by $N (v_\ell)$.
The factor associated with $c_p$ and derived from parity equation \eqref{equation:ParityConstraint} can be expressed as an indicator function
\begin{equation} \label{equation:LocalFactors}
\mathcal{G}_{p} ( \vv_{p} ) = \mathbf{1} \left( \sum_{v_\ell \in N(c_p)} \omega_{\ell, p} \otimes v_{\ell} = 0 \right)
\end{equation}
where $\vv_{p} = \left( v_{\ell} \in N(c_p) \right)$ is a shorthand notation for the restriction of $\vv$ to entries associated with graph neighborhood $N(c_p)$.
With these definitions, the factor function associated with our LDPC code assumes the product decomposition given by
\begin{equation} \label{equation:FactorFunction}
\begin{split}
\mathcal{G} ( \vv )
= \prod_{p \in [L(1-R)]} \mathcal{G}_{p} ( \vv_{p} ) .
\end{split}
\end{equation}
In words, $\mathcal{G} ( \vv )$ is an indicator function that assesses whether its argument is a valid codeword.

\subsection{Belief Propagation}

We view messages for a non-binary LDPC code as multi-dimensional belief vectors over $\mathbb{F}_q$.
Messages from variable nodes to check nodes are denoted by $\boldsymbol{\mu}_{v \to c}$, and messages in the reverse direction are represented by $\boldsymbol{\mu}_{c \to v}$.
Formally, a message going from check node $c_p$ to variable node $v_{\ell} \in N(c_p)$ is computed component-wise through
\begin{equation} \label{equation:BP-Check2Variable}
\boldsymbol{\mu}_{c_p \to v_{\ell}} (g)
= \sum_{\vv_{p}: v_{\ell} = g} \mathcal{G}_{p} \left( \vv_{p} \right)
\prod_{v_j \in N(c_p) \setminus v_{\ell}} \boldsymbol{\mu}_{v_j \to c_p} (g_j) .
\end{equation}
While \eqref{equation:BP-Check2Variable} is shown in compact form, the actual summation operation is cumbersome.
Finding the set of summands entails identifying sequences of the form $\left( g_j \in \mathbb{F}_q: v_j \in N(c_p) \setminus v_{\ell} \right)$ that fulfill local condition \eqref{equation:ParityConstraint} or, equivalently,
\begin{equation} \label{equation:LocalCondition}
\sum_{v_j \in N(c_p) \setminus v_{\ell}} \omega_{j, p} \otimes g_j
= - \omega_{\ell, p} \otimes g .
\end{equation}
A belief vector passed from variable node $v_{\ell}$ to check node $c_p$, where $p \in N(v_{\ell})$, is calculated component-wise via
\begin{equation} \label{equation:BP-Variable2Check}
\boldsymbol{\mu}_{v_{\ell} \rightarrow c_p} (g)
\propto \boldsymbol{\alpha}_{\ell} (g) \prod_{c_{\xi} \in N(v_{\ell}) \setminus c_p} \boldsymbol{\mu}_{c_{\xi} \to v_{\ell}}(g) .
\end{equation}
The `$\propto$' symbol indicates that the positive measure can be normalized before being sent out as a message.
Vector $\boldsymbol{\alpha}_{\ell}$ in \eqref{equation:BP-Variable2Check} can be viewed as a collection of beliefs based on local observations, as in \eqref{equation:ConditionalProbability}.
That is, entry $\boldsymbol{\alpha}_{\ell}(g)$ captures the probability that symbol $g$ is the true field element within section~$\ell$.
Other BP messages are initialized with $\boldsymbol{\mu}_{v \to c} = \mathbf{1}$ and $\boldsymbol{\mu}_{c \to v} = \mathbf{1}$.
% These message passing operations appear in Fig.~\ref{figure:FactorGraph}.
The traditional parallel sum-product algorithm iterates between \eqref{equation:BP-Check2Variable} and \eqref{equation:BP-Variable2Check}, alternating between updated rightbound messages and leftbound messages.

One of the key advantages of indexing vectors using field elements in $\mathbb{F}_q$, as pointed out in \cite{bennatan2006design}, is the ensuing ability to define pertinent operators on these vectors.
Paralleling established literature, we define two actions.

\begin{definition}[Vector $+g$ Operator \cite{bennatan2006design}]
For field element $g \in \mathbb{F}_q$, the vector $+g$ operator acting on $\bv$ and denoted by $\bv^{+g}$ is defined as
\begin{equation*}
\begin{split}
\bv^{+g} &= \left( b_g, b_{g \oplus 1}, \ldots, b_{g \oplus (q-1)} \right) \\
&= \left( b_{h \oplus g} : h \in \mathbb{F}_q \right)
\end{split}
\end{equation*}
where subscript addition $\oplus$ is performed in $\mathbb{F}_q$.
\end{definition}

\begin{definition}[Vector $\times g$ Operator \cite{bennatan2006design}] \label{defintion:TimesOperator}
For field element $g \in \mathbb{F}_g \setminus \{ 0 \}$, we define the vector $\times g$ operator acting on $\bv$ and denoted by $\bv^{\times g}$ by
\begin{equation*}
\begin{split}
\bv^{\times g} &= \left( b_0, b_g, b_{2 \otimes g}, \ldots, b_{(q-1) \otimes g} \right) \\
&= \left( b_{h \otimes g} : h \in \mathbb{F}_q \right)
\end{split}
\end{equation*}
where subscript product $\otimes$ takes place in $\mathbb{F}_q$.
\end{definition}

% We stress that the $+g$ and $\times g$ operators introduced above are reversible, with
% \begin{xalignat*}{2}
% \left( \bv^{+g} \right)^{- g}
% &= \bv & g &\in \mathbb{F}_q \\
% \left( \bv^{\times g} \right)^{\times g^{-1}} &= \bv
% & g &\in \mathbb{F}_q \setminus \{ 0 \} .
% \end{xalignat*}
% They essentially permute the entries of $\bv$ in a structured fashion that naturally meshes with field actions.
% These operators can be applied to probability vector on $\mathbb{F}_q$, and they are especially meaningful in the computation of BP messages for non-binary LDPC codes.
% Within the $\mathbb{F}_q$ LPDC graph, factor nodes impose constraints that are easily expressible within Galois field $\mathbb{F}_q$, as in \eqref{equation:ParityConstraint}.
% Vector operators then become a convenient way to track the distribution of belief vectors during message passing.
% Specifically, under these operations, we can rewrite \eqref{equation:BP-Check2Variable} in a concise manner:
Under these operations, we can rewrite \eqref{equation:BP-Check2Variable} in a concise manner:
\begin{equation} \label{equation:FactorOperationField1}
\boldsymbol{\mu}_{c_p \to v_{\ell}}
= \left( \bigodot_{v_j \in N(c_p) \setminus v_{\ell}} \left( \boldsymbol{\mu}_{v_j \to c_p} \right)^{\times \omega_{j,p}^{-1}} \right)^{\times \left(- \omega_{\ell,p} \right)}
\end{equation}
where $\omega_{j,p}$ is the label on the edge between variable node $v_j$ and factor node $c_p$.
The operator $\odot$ denotes the $\mathbb{F}_q$ convolution between two vectors,
\begin{equation*}
\left[ \boldsymbol{\mu} \odot \boldsymbol{\nu} \right]_{g}
= \sum_{h \in \mathbb{F}_q} \mu_h \cdot \nu_{g - h}
\qquad g \in \mathbb{F}_q .
\end{equation*}
The exposition can be simplified further if we absorb the edge labels within the messages themselves.
Specifically, we adopt the definitions
\begin{align}
\overline{\boldsymbol{\mu}}_{v_j \to c_p} &= \left( \boldsymbol{\mu}_{v_j \to c_p} \right)^{\times \omega_{j,p}^{-1}}
\label{equation:AbsorbedMessageVar2Check} \\
\overline{\boldsymbol{\mu}}_{c_p \to v_{\ell}} &= \left( \boldsymbol{\mu}_{c_p \to v_{\ell}} \right)^{\times \left(- \omega_{\ell,p}^{-1} \right)} .
\label{equation:AbsorbedMessageCheck2Var}
\end{align}
Then \eqref{equation:FactorOperationField1} morphs into the simpler expression
\begin{equation} \label{equation:BP-Check2Variable-Convolution}
\overline{\boldsymbol{\mu}}_{c_p \to v_{\ell}}
= \bigodot_{v_j \in N(c_p) \setminus v_{\ell}} \overline{\boldsymbol{\mu}}_{v_j \to c_p} .
\end{equation}
This equation highlights the role of the $\mathbb{F}_q$ convolution within BP for non-binary LDPC codes.
The message from variable node $v_{\ell}$ to check node $c_p$ found in \eqref{equation:BP-Variable2Check} also admits a more compact form.
For $p \in N(v_{\ell})$, the traditional outgoing message from a variable node can be written as
\begin{equation} \label{equation:VariableNodeEstimate}
\boldsymbol{\mu}_{v_{\ell} \to c_p}
= \frac{ \boldsymbol{\alpha}_{\ell} \circ \left( \operatorname*{\bigcirc}_{c_{\xi} \in N(v_{\ell}) \setminus c_p} \boldsymbol{\mu}_{c_{\xi} \to v_{\ell}} \right) }
{ \left\| \boldsymbol{\alpha}_{\ell} \circ \left( \operatorname*{\bigcirc}_{c_{\xi} \in N(v_{\ell}) \setminus c_p} \boldsymbol{\mu}_{c_{\xi} \to v_{\ell}} \right) \right\|_1 }
\end{equation}
where $\circ$ represents the Hadamard product.

% At any point during the iterative process, the belief vector associated with $v_{\ell}$ and derived from extrinsic information is proportional to the element-wise product of the incoming messages from adjacent parity factors,
% % CHECK
% \begin{equation} \label{equation:EquivalentPriors}
% \boldsymbol{\mu}_{v_{\ell}}(g) = \frac{\prod_{c_p \in N(v_{\ell})} \boldsymbol{\mu}_{c_p \to v_{\ell}} (g)}
% { \left\| \operatorname*{\bigcirc}_{c_p \in N(v_{\ell})} \boldsymbol{\mu}_{c_p \to v_{\ell}} \right\|_1 } .
% \end{equation}
A natural estimate for the distribution associated with variable node~$v_{\ell}$, including intrinsic information, is
\begin{equation} \label{equation:BP-Estimate}
\begin{split}
\hat{\Sv}_{\ell}
&= \frac{ \boldsymbol{\alpha}_{\ell} \circ \left( \operatorname*{\bigcirc}_{c_{p} \in N(v_{\ell})} \boldsymbol{\mu}_{c_{p} \to v_{\ell}} \right) }
{ \left\| \boldsymbol{\alpha}_{\ell} \circ \left( \operatorname*{\bigcirc}_{c_{p} \in N(v_{\ell})} \boldsymbol{\mu}_{c_{p} \to v_{\ell}} \right) \right\|_1 }.
\end{split}
\end{equation}
As we will shortly see, \eqref{equation:BP-Estimate} is the output of our proposed denoiser. 

\begin{remark}
\label{remark:fwht}
In our construction, $q = 2^m, m \geq 1$ because indexing is derived from sequences of bits.
This invites the application of fast techniques to implement message passing over the corresponding factor graph.
Specifically, the fast Walsh-Hadamard transform (FWHT) can be utilized to rapidly and efficiently compute \eqref{equation:BP-Check2Variable-Convolution}, with
\begin{equation*}
\overline{\boldsymbol{\mu}}_{c_p \to v_{\ell}}
\propto \operatorname{fwht}^{-1} \left( \prod_{v_j \in N(c_p) \setminus v_{\ell}} \operatorname{fwht} \left( \overline{\boldsymbol{\mu}}_{v_j \to c_p} \right) \right) .
\end{equation*}
% This technique is especially meaningful given that it may be desirable to maintain large sections and, hence, a large alphabet size for sparse regression codewords.
Alternatively, one could adopt a different finite field convolution or a ring structure amenable to the circular convolution to create local factor functions conducive to the fast Fourier transform \cite{davey1998low,song2003reduced,goupil2007fft}.
\end{remark}

With these tools in mind, we are ready to formally define our proposed denoiser. 

\subsection{Sparse Regression LDPC Denoiser}

Conceptually, one can initiate the state of the LDPC factor graph using the effective observation $\rv$, run BP, and then gain an estimate for the state based on \eqref{equation:BP-Estimate}.
% This procedure offers a blueprint for denoising.
As mentioned before, in the absence of BP iterations, local estimates reduce to the conditional expectation $\mathbb{E} \left[ \Sv_{\ell} \middle| \Rv_{\ell} = \rv_{\ell} \right]$ found in \eqref{equation:LocalMMSE}.
Yet, as more iterations of the BP algorithm are performed, the estimate for $\Sv_{\ell}$ is refined based on the computation tree of the outer code, up to a certain depth.

\begin{definition}[BP-N Denoiser] \label{definition:BPDenoiser}
Let $N$ denote the number of BP iterations to perform, where $N$ is less than the girth of the LDPC factor graph. 
The BP-N denoiser
\begin{enumerate}
    \item initializes the LDPC factor graph with estimates $\boldsymbol{\alpha}_{\ell}$ computed from $\rv_{\ell}$ for $\ell \in [L]$ according to \eqref{equation:ConditionalProbability};
    \item computes and passes variable to check messages (see \eqref{equation:VariableNodeEstimate}) and check to variable messages (see \eqref{equation:AbsorbedMessageVar2Check}, \eqref{equation:BP-Check2Variable-Convolution}, \eqref{equation:AbsorbedMessageCheck2Var} and Remark~\ref{remark:fwht}) along the edges of the factor graph in an alternating fashion $N$ times;
    \item computes updated state estimates according to \eqref{equation:BP-Estimate}.
\end{enumerate}
The output of this denoiser can then be passed to the AMP composite algorithm for the computation of the next residual, enhanced with the Onsager term.
\end{definition}

The derivation of the Onsager correction term corresponding to this denoiser is included in Appendix~\ref{appendix:OnsagerDerivation}. 
At this point, we turn to a performance assessment of sparse regression LDPC codes via numerical simulations.

\section{Simulation Results}
\label{section:SimulationResults}
In this section, we seek to characterize the performance of SR-LDPC codes.\footnote{The source code used to generate these results is available online at https://github.com/EngProjects/mMTC/tree/code.}
As benchmarks, we compare our results to the concatenated SPARC/LDPC scheme presented in \cite{greig2017techniques} as well as 5G-NR binary LDPC encoded BPSK. 
We note that the latter comparison provides limited insight because the SR-LDPC code is allowed real channel inputs whereas the 5G-NR code is constrained to binary inputs; nevertheless, we make the comparison to see whether SR-LDPC codes are competitive with state of the art, commonly used codes. 

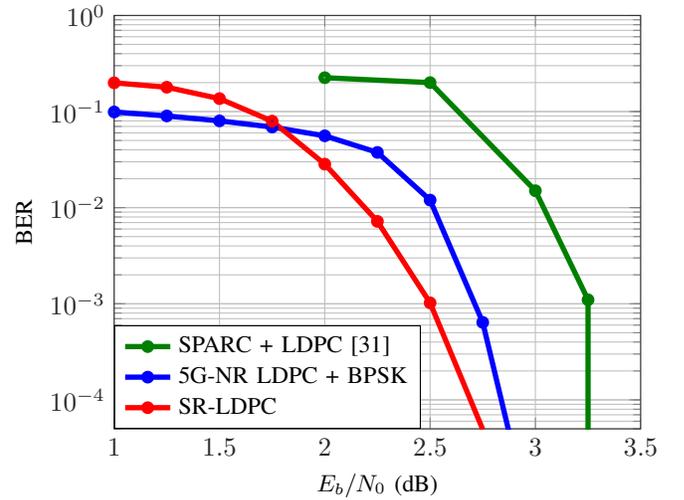
\begin{figure}[t]
    \centering
    \begin{tikzpicture}

\definecolor{customred}{rgb}{0.63529,0.07843,0.18431} % red
\definecolor{customblue}{rgb}{0.00000,0.44706,0.74118} % blue
\definecolor{customgreen}{rgb}{0.00000,0.49804,0.00000} % dark green

\begin{semilogyaxis}[
    font=\small,
    width=7cm,
    height=5.5cm,
    scale only axis,
    every outer x axis line/.append style={white!15!black},
    every x tick label/.append style={font=\color{white!15!black}},
    xmin=1,
    xmax=3.5,
    xtick = {1.0, 1.5, 2.0, 2.5, 3.0, 3.5},
    xlabel={$E_b/N_0$ (dB)},
    xmajorgrids,
    every outer y axis line/.append style={white!15!black},
    every y tick label/.append style={font=\color{white!15!black}},
    ymin=0.00005,
    ymax=1,
    ytick = {0.00001, 0.0001, 0.001, 0.01, 0.1, 1.0},
    ylabel={BER},
    ymajorgrids,
    yminorgrids,
    legend style={at={(0,0)},anchor=south west, draw=black,fill=white,legend cell align=left}
]

\addplot [
    color=customgreen,
    solid,
    line width=2.0pt,
    mark size=1.4pt,
    mark=o,
    mark options={solid}
]
table[row sep=crcr]{
    2.0 2.25e-1 \\
    2.5 2e-1 \\
    3.0 1.5e-2 \\
    3.25 1.1e-3 \\
    3.25 0.0000001 \\
    3.5 0 \\
    3.75 0 \\
};
\addlegendentry{SPARC + LDPC \cite{greig2017techniques}};

\addplot [
    color=blue,
    solid,
    line width=2.0pt,
    mark size=1.4pt,
    mark=o,
    mark options={solid}
]
table[row sep=crcr]{
    1.0 0.099 \\
    1.25 0.09 \\
    1.5 0.08 \\
    1.75 0.069 \\
    2.0 0.0559 \\
    2.25 0.03757 \\
    2.5 0.011984 \\
    2.75 0.00063983 \\
    3.0 0.0000033062 \\
};
\addlegendentry{5G-NR LDPC + BPSK};

\addplot [
    color=red,
    solid,
    line width=2.0pt,
    mark size=1.4pt,
    mark=o,
    mark options={solid}
]
table[row sep=crcr]{
    1.0 0.1991 \\
    1.25 0.1791 \\
    1.5 0.1363 \\
    1.75 0.0795 \\
    2.0 0.028355 \\
    2.25 0.0072 \\
    2.5 0.001022 \\
    2.75 0.00004927\\
    % 3.0 3.22e-6 \\
};
\addlegendentry{SR-LDPC};

\end{semilogyaxis}
\end{tikzpicture}
    \caption{BER performance comparison of a sparse regression LDPC code, the concatenated SPARC/LDPC construction from \cite{greig2017techniques}, and 5G-NR binary LDPC encoded BPSK as a function of $E_b/N_0$. The SR-LDPC code requires a lower $E_{\mathrm{b}}/N_0$ to achieve a target BER than LDPC encoded BPSK and the scheme in \cite{greig2017techniques}. }
    \label{fig:isit2023results:ber_comparison}
\end{figure}

The scenario of interest is one in which $5888$ information bits are to be transmitted over $7350$ real channel uses. 
We define the SR-LDPC rate to be the number of information bits over the number of channel uses; thus, we have that $R_{\mathrm{SRLDPC}} \approx 0.80$. 
The non-binary LDPC code employed is a $(751, 736)$ code ($R_{\mathrm{LDPC}} \approx 0.98$) over GF($256$) whose edges are generated via progressive edge growth (PEG) and whose weights were chosen uniformly at random from the elements of $\mathbb{F}_{256}\setminus 0$. 
The fraction of degree-$2$ variable nodes is chosen to be higher than usual so that the AMP-BP iterations can bootstrap in high-noise environments. 
The AMP-BP decoder is run for up to $100$ iterations and then the non-binary LDPC BP decoder is run for another $100$ iterations, or until a valid codeword is obtained. 
Figure~\ref{fig:isit2023results:ber_comparison} highlights our results. 

From Fig.~\ref{fig:isit2023results:ber_comparison}, it is clear that sparse regression LDPC codes provide a roughly $0.75$~dB improvement at a BER of $10^{-3}$ over other SPARC/LDPC concatenated coding structures. 
Furthermore, the SR-LDPC code outperforms 5G-NR LDPC encoded BPSK in some regimes; we note however, that 5G-NR LDPC encoded BPSK has a lower error floor than the SR-LDPC code. 

\section{Conclusion}
\label{section:conclusion}

In conclusion, this article presents sparse regression LDPC codes and their decoding. 
A sparse regression code is a concatenated structure with an inner SPARC-like code and an outer non-binary LDPC code. 
The inner code is decoded using AMP with a dynamic denoiser which runs BP on the factor graph of the outer LDPC code to improve the state estimate at every iteration.
After running many iterations of the AMP-BP decoder, the final effective observation is fed into a standard LDPC BP decoder, which seeks to correct any residual errors that may be present in the received signal. 
Numerical results show that sparse regression codes exhibit performance improvements over standard LDPC codes and other concatenated SPARC/LDPC constructions. 

As mentioned in the introduction, both SPARCs and LDPC codes are known to achieve capacity in certain circumstances, yet both codes generally suffer at short block lengths. 
The success of sparse regression LDPC codes is evidence that SPARCs and LDPC codes can be seamlessly combined to significantly improve performance.
Directions for future work include optimizing the outer LDPC code and applying coded demixing \cite{ebert2022coded} to SR-LDPC codes to accommodate multiple users.
% Add sentence about future work - outer LDPC code, power allocation, MU-SR-LDPC, etc
% This observation is further explored in another paper to be published alongside this article which introduces a theoretical framework for tracking the MSE through the AMP-BP iteration. 

\clearpage
\bibliographystyle{IEEEbib}
\bibliography{IEEEabrv,isit_results}
\clearpage

\appendices

\section{Derivation of Onsager Term}
\label{appendix:OnsagerDerivation}

Intuitively, the role of the Onsager term is to (asymptotically) cancel the first-order correlations between $\Am^{\mathrm{T}} \zv^{(t)}$ and $\sv^{(t)}$ and thereby maintain a structure conducive to prompt convergence and analysis.
This factor, emblematic of AMP algorithms, appears in \eqref{equation:AMP-Residual} and is given by
\begin{equation*}
\begin{split}
&\frac{\zv^{(t-1)}}{n} \operatorname{div} \etav_{t-1} \left( \Am^{\mathrm{T}} \zv^{(t-1)} + \sv^{(t-1)} \right) \\
&= \frac{\zv^{(t-1)}}{n} \operatorname{div} \etav_{t-1} \left( \rv^{(t-1)} \right)
\end{split}
\end{equation*}
where the $\operatorname{div}$ operator can be expanded into
\begin{equation} \label{equation:DivOperator}
\operatorname{div} \etav \left( \rv \right)
= \sum_{\ell \in [L]} \operatorname{div} \hat{\sv}_{\ell} ( \rv, \tau )
= \sum_{\ell \in [L]} \sum_{g \in \mathbb{F}_q} \frac{\partial \hat{s}_{\ell} \left( g, \rv, \tau \right)}{\partial \rv_{\ell}(g)} .
\end{equation}
Thus, as an intermediate step, we must calculate the partial derivative of $\hat{s}_{\ell} \left( g, \rv, \tau \right)$ with respect to $\rv_{\ell}(g)$.
This computation is rendered much simpler under the following condition. 
% A situation that makes this computation simpler is the following condition.

\begin{condition}[Sub-Girth AMP-BP] \label{condition:Sub-Girth AMP-BP}
The BP-N denoiser is said to possess the \emph{sub-girth AMP-BP} condition when fewer message passing iterations are performed on the factor graph of the LDPC code than the shortest cycle of this same graph, per AMP denoising step.
\end{condition}

This condition ensures that the message passing operations employed during denoising yield valid computation trees without cycles.
Fortunately, this is the regime we are primarily interested in.

\begin{lemma} \label{lemma:BP-PartialSestimate}
Under Condition~\ref{condition:Sub-Girth AMP-BP}, the partial derivative of $\hat{s}_{\ell} \left( g, \rv, \tau \right)$ with respect to $\rv_{\ell}(g)$ is equal to
\begin{equation} \label{equation:BP-PartialSestimate}
\frac{\partial \hat{s}_{\ell} \left( g, \rv, \tau \right)}{\partial \rv_{\ell}(g)}
= \frac{1}{\tau^2} \hat{s}_{\ell} \left( g, \rv, \tau \right)
\left( 1 - \hat{s}_{\ell} \left( g, \rv, \tau \right) \right)
\end{equation}
where $g \in \mathbb{F}_q$.
\end{lemma}
\begin{IEEEproof}
Recall that the output of the BP denoiser defined in \eqref{equation:BP-Estimate} can be expressed as
\begin{equation}
\begin{split}
\hat{s}_{\ell} \left( g, \rv, \tau \right)
&= \frac{ \boldsymbol{\alpha}_{\ell} (g) \prod_{p \in N(v_{\ell})} \boldsymbol{\mu}_{c_{p} \to v_{\ell}}(g) }
{ \sum_{h \in \mathbb{F}_q} \boldsymbol{\alpha}_{\ell} (h) \prod_{p \in N(v_{\ell})} \boldsymbol{\mu}_{c_{p} \to v_{\ell}}(h) } \\
&= \frac{ \boldsymbol{\alpha}_{\ell} (g) \boldsymbol{\mu}_{v_{\ell} \to c_0}(g) }
{ \sum_{h \in \mathbb{F}_q} \boldsymbol{\alpha}_{\ell} (h) \boldsymbol{\mu}_{v_{\ell} \to c_0}(h) } .
\end{split}
\end{equation}
Under Condition~\ref{condition:Sub-Girth AMP-BP}, belief vector $\boldsymbol{\mu}_{v_{\ell} \to c_0}$ is based solely on extrinsic information and, hence, it is determined based on $\left\{ \rv_j : j \in [L] \setminus \{ \ell \} \right\}$.
Consequence, we gather that
\begin{equation*}
\frac{\partial \boldsymbol{\mu}_{v_{\ell} \to c_0}(g)}{\partial \rv_{\ell}(g)} = 0 .
\end{equation*}
Under such circumstances, we can calculate the desired derivative in a straightforward manner, with
\begin{equation*}
\begin{split}
\frac{\partial \hat{s}_{\ell} \left( g, \rv, \tau \right)}{\partial \rv_{\ell}(g)}
% &= \frac{\partial}{\partial \rv_{\ell}(g)}
% \frac{ \boldsymbol{\alpha}_{\ell} (g) \boldsymbol{\mu}_{v_{\ell} \to c_0}(g) }
% { \sum_{h \in \mathbb{F}_q} \boldsymbol{\alpha}_{\ell} (h) \boldsymbol{\mu}_{v_{\ell} \to c_0}(h) } \\
&= \frac{\partial}{\partial \rv_{\ell}(g)}
\frac{ e^{\frac{ \rv_{\ell}(g)}{\tau^2}} \boldsymbol{\mu}_{v_{\ell} \to c_0}(g) }
{ \sum_{h \in \mathbb{F}_q} e^{\frac{ \rv_{\ell}(h)}{\tau^2}} \boldsymbol{\mu}_{v_{\ell} \to c_0}(h) } \\
&= \frac{1}{\tau^2} \frac{ e^{\frac{ \rv_{\ell}(g)}{\tau^2}} \boldsymbol{\mu}_{v_{\ell} \to c_0}(g) }{\sum_{h \in \mathbb{F}_{q}} e^{\frac{ \rv_{\ell}(g)}{\tau^2}} \boldsymbol{\mu}_{v_{\ell} \to c_0}(g) } \\
&\quad- \frac{1}{\tau^2}\frac{ \left( e^{\frac{ \rv_{\ell}(g)}{\tau^2}} \boldsymbol{\mu}_{v_{\ell} \to c_0}(g) \right)^2 }{\left( \sum_{h \in \mathbb{F}_{q}} e^{\frac{ \rv_{\ell}(g)}{\tau^2}} \boldsymbol{\mu}_{v_{\ell} \to c_0}(g) \right)^2} \\
&= \frac{1}{\tau^2} \hat{s}_{\ell} \left( g, \rv, \tau \right)
\left( 1 - \hat{s}_{\ell} \left( g, \rv, \tau \right) \right) .
\end{split}
\end{equation*}
This last line corresponds to the statement of the lemma.
\end{IEEEproof}

It is worth emphasizing that the derivative in \eqref{equation:BP-PartialSestimate} remains unchanged irrespective of the number of BP rounds computed on the factor graph, so long as Condition~\ref{condition:Sub-Girth AMP-BP} is satisfied.
The divergence of \eqref{equation:DivOperator} assumes the same simple form under such circumstances.

\begin{proposition} \label{proposition:DivComputationBP}
The divergence of $\etav \left( \rv \right)$ with respect to $\rv$ is equal to
\begin{equation} \label{equation:BP-OnsagerCorrection}
\begin{split}
&\operatorname{div} \etav \left( \rv \right)
= \frac{1}{\tau^2} \left( \left\| \etav \left( \rv \right) \right\|_1 - \left\| \etav \left( \rv \right) \right\|^2 \right) .
\end{split}
\end{equation}
\end{proposition}
\begin{IEEEproof}
We expand the $\operatorname{div}$ operator as
\begin{equation} \label{equation:OnsagerDerivation}
\begin{split}
\operatorname{div} \etav \left( \rv \right)
&= \sum_{\ell \in [L]} \operatorname{div} \hat{\sv}_{\ell} \left( \rv, \tau \right)
= \sum_{\ell \in [L]} \sum_{g \in \mathbb{F}_q} \frac{\partial \hat{s}_{\ell} \left( g, \rv, \tau \right)}{\partial \rv_{\ell}(g)} \\
&= \sum_{\ell \in [L]} \sum_{g \in \mathbb{F}_q} \frac{1}{\tau^2} \hat{s}_{\ell} \left( g, \rv, \tau \right)
\left( 1 - \hat{s}_{\ell} \left( g, \rv, \tau \right) \right) \\
&= \frac{1}{\tau^2} \left( \left\| \etav \left( \rv \right) \right\|_1
- \left\| \etav \left( \rv \right) \right\|^2 \right) .
\end{split}
\end{equation}
The last equality follows from the fact that, since $\hat{s}_{\ell} \left( g, \rv, \tau \right)$ lies between zero and one, the corresponding partial derivative with respect to $\rv_{\ell}(g)$ found in Lemma~\ref{lemma:BP-PartialSestimate} is always non-negative.
\end{IEEEproof}

\section{Denoiser is Lipschitz Continuous}
\label{appendix:LipschitzDenoiser}

The mathematical underpinnings for the rigorous application of AMP in this article are presented by Berthier, Montanari, and Nguyen in~\cite{berthier2020state}.
One of the conditions for the state evolution to hold for non-separable functions is that the denoiser must be pseudo-Lipschitz of a certain order.
For the problem at hand, we are able to show the stronger Lipschitz condition, which is sufficient.
Thus, the main objective of this section is to demonstrate that the denoiser introduced in Definition~\ref{definition:BPDenoiser} is Lipschitz continuous under Condition~\ref{condition:Sub-Girth AMP-BP}.
% Since we are primarily interested in denoising functions that perform a few rounds of message passing on the outer LDPC factor graph, this condition is reasonable.
To achieve this goal, our strategy is to demonstrate that the magnitudes of the entries in the Jacobian matrix of $\etav(\rv)$ with respect to $\rv$ are uniformly bounded.
% The derivations below are somewhat tedious, which explains their location in an appendix.

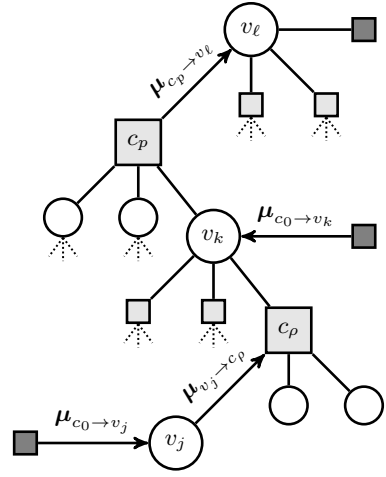
\begin{figure}[t]
  \centering
  \begin{tikzpicture}
  [
  font=\small, line width=1pt, >=stealth', draw=black,
  check/.style={rectangle, minimum height=6mm, minimum width=6mm, draw=black, fill=gray!20},
  smallcheck/.style={rectangle, minimum height=3mm, minimum width=3mm, draw=black, fill=gray!20},
  trivialcheck/.style={rectangle, minimum height=3mm, minimum width=3mm, draw=black, fill=gray},
  section/.style={circle, minimum size=7mm, draw=black},
  emptysection/.style={circle, minimum size=5mm, draw=black}
  ]

\tikzset{
  connect/.pic={
    \draw[densely dotted, line width=0.75pt] (0,0) to (0,-0.3125);
    \draw[densely dotted, line width=0.75pt] (0,0) to (-0.1875,-0.25);
    \draw[densely dotted, line width=0.75pt] (0,0) to (0.1875,-0.25);
  }
}

\node[section] (v0) at (0.5,0.25) {$v_{\ell}$};
\node[trivialcheck] (t0) at (2,0.25) {}
  edge (v0);
\node[check] (c0a) at (-1,-1.25) {$c_p$};
\draw[->] (c0a) -- node[above,rotate=45]{$\boldsymbol{\mu}_{c_p \to v_{\ell}}$} (v0);
\node[smallcheck] (c0b) at (0.5,-0.75) {}
  edge[-] (v0);
\pic at (c0b.south) {connect};
\node[smallcheck] (c0c) at (1.5,-0.75) {}
  edge[-] (v0);
\pic at (c0c.south) {connect};

\node[emptysection] (v1a) at (-2,-2.25) {}
  edge[-] (c0a);
\pic at (v1a.south) {connect};
\node[emptysection] (v1b) at (-1,-2.25) {}
  edge[-] (c0a);
\pic at (v1b.south) {connect};
\node[section] (v1c) at (0,-2.5) {$v_k$}
  edge[-] (c0a);
\node[trivialcheck] (t1c) at (2,-2.5) {};
\draw[->] (t1c) -- node[above]{$\boldsymbol{\mu}_{c_0 \to v_k}$} (v1c);

\node[smallcheck] (c1a) at (-1,-3.5) {}
  edge[-] (v1c);
\pic at (c1a.south) {connect};
\node[smallcheck] (c1b) at (0,-3.5) {}
  edge[-] (v1c);
\pic at (c1b.south) {connect};
\node[check] (c1c) at (1,-3.75) {$c_{\rho}$}
  edge[-] (v1c);

\node[section] (v2a) at (-0.5,-5.25) {$v_j$};
\draw[->] (v2a) -- node[above,rotate=45]{$\boldsymbol{\mu}_{v_j \to c_{\rho}}$} (c1c);
\node[emptysection] (v2b) at (1,-4.75) {}
  edge[-] (c1c);
\node[emptysection] (v2c) at (2,-4.75) {}
  edge[-] (c1c);
\node[trivialcheck] (t1c) at (-2.5,-5.25) {};
\draw[->] (t1c) -- node[above]{$\boldsymbol{\mu}_{c_0 \to v_j}$} (v2a);

\end{tikzpicture}
  \caption{Computation tree for variable node $v_{\ell}$ obtained by taking the factor graph of the outer LDPC code, setting $v_{\ell}$ as the root, and retaining only the nodes involved in the computation of $\hat{\sv}_{\ell}\left(\rv, g\right)$. We use this data structure to compute the derivatives in \eqref{equation:JacobianMatrixEntries}. }
  \label{figure:computation_tree_illustration}
\end{figure}

Recall that the denoiser assumes a sectional form, as described in Definition~\ref{definition:BPDenoiser}.
The vector estimate for section~$\ell$ becomes
\begin{equation*}
\hat{\sv}_{\ell} (\rv)
% = \sum_{g \in \mathbb{F}_q} \hat{\vv}_{\ell} (g) \ev_g
= \sum_{g \in \mathbb{F}_q} \Pr \left( V_{\ell} = g \middle| \Rv_{\mathrm{tree}} = \rv_{\mathrm{tree}} \right) \ev_g ,
\end{equation*}
where $\Rv_{\mathrm{tree}}$ denotes the measurements associated with the computational tree of the LDPC code rooted at section~$\ell$.
The (realized) scaling factors found in \eqref{equation:BP-Estimate} are given by
% JE NOTATION: N_0 is set of neighbors including local observation
\begin{equation}
\begin{split}
\hat{\sv}_{\ell} (\rv)
% &= \frac{ \bar{\boldsymbol{\alpha}}_{\ell} \circ \left( \operatorname*{\bigcirc}_{p \in N(v_{\ell})} \boldsymbol{\mu}_{c_p \to v_{\ell}} \right) }
% { \left\| \bar{\boldsymbol{\alpha}}_{\ell} \circ \left( \operatorname*{\bigcirc}_{p \in N(v_{\ell})} \boldsymbol{\mu}_{c_p \to v_{\ell}} \right) \right\|_1 } \\
&= \frac{ \operatorname*{\bigcirc}_{p \in N_0(v_{\ell})} \boldsymbol{\mu}_{c_{p} \to v_{\ell}} }
{ \left\| \operatorname*{\bigcirc}_{p \in N_0(v_{\ell})} \boldsymbol{\mu}_{c_{p} \to v_{\ell}} \right\|_1 } .
\end{split}
\end{equation}
where $N_0(v_{\ell})$ denotes the neighborhood of $v_{\ell}$ including the local observation and $\boldsymbol{\mu}_{c_0 \to v_{\ell}} = \boldsymbol{\alpha}_{\ell}$.
% We point out that $\mathbb{F}_q$ vector indexing is employed in these expressions.
We are ultimately interested in Jacobian entries of the form
\begin{equation} \label{equation:JacobianMatrixEntries}
\begin{split}
&\frac{\partial \hat{\sv}_{\ell} \left( \rv, g \right)}{\partial \rv_j(h)}
= \frac{\partial }{\partial \rv_j(h)}
\frac{ \prod_{p \in N_0(v_{\ell})} \boldsymbol{\mu}_{c_{p} \to v_{\ell}} (g) }
{ \left\| \operatorname*{\bigcirc}_{p \in N_0(v_{\ell})} \boldsymbol{\mu}_{c_{p} \to v_{\ell}} \right\|_1 } \\
&= \frac{\partial }{\partial \rv_j(h)}
\frac{ \boldsymbol{\alpha}_{\ell}(g) \prod_{p \in N(v_{\ell})} \boldsymbol{\mu}_{c_{p} \to v_{\ell}} (g) }
{ \sum_{k \in \mathbb{F}_q} \boldsymbol{\alpha}_{\ell}(k) \prod_{p \in N(v_{\ell})} \boldsymbol{\mu}_{c_{p} \to v_{\ell}} (k) }
\end{split}
\end{equation}
for $g, h \in \mathbb{F}_q$ and $\ell, j \in [L]$.
We adopt a divide-and-conquer approach to identify and bound these derivatives.
Specifically, we focus on the rooted tree obtained by taking the factor graph of the outer LDPC code, setting $v_{\ell}$ as the root, and retaining only the nodes involved in the computation of $\hat{\sv}_{\ell} ( \rv, g )$. 
Under Condition~\ref{condition:Sub-Girth AMP-BP}, this sub-graph must form a proper tree; Fig.~\ref{figure:computation_tree_illustration} offers a notional diagram to illustrate the outcome of this process.

We seek to bound the magnitude of the derivatives in \eqref{equation:JacobianMatrixEntries} based on the distance between $v_{\ell}$ and $v_j$ in this rooted tree.
We begin with local observations.

\begin{proposition} (Local Observations) \label{proposition:DerivativeAlphaEll}
The partial derivatives of $\boldsymbol{\alpha}_j$ with respect to $\rv_k(h)$ are given by
\begin{equation}
\label{equation:DerivativeAlpha}
\frac{\partial \boldsymbol{\alpha}_j}{\partial \rv_k(h)} = 
\begin{cases}
\frac{1}{\tau^2} \boldsymbol{\alpha}_j (h) \left( \ev_h - \boldsymbol{\alpha}_j \right) & j = k \\
0 & j \neq k \\
\end{cases}
\end{equation}
$\forall h \in \mathbb{F}_{q}$ and where $\tau > 0$ is the standard deviation of the effective observation.
\end{proposition}
\begin{IEEEproof}
As defined in \eqref{equation:ConditionalProbability}, the vector $\boldsymbol{\alpha}_j$ is given by
\begin{equation*}
\boldsymbol{\alpha}_j (g)
= \frac{ e^{\frac{ \rv_j(g) }{\tau^2}} }{\sum_{k \in \mathbb{F}_q} e^{\frac{ \rv_j(k) }{\tau^2}}}
\qquad \forall g \in \mathbb{F}_q .
\end{equation*}
When $j \neq k$, it immediately follows that $\partial\boldsymbol{\alpha}_j/\partial\rv_k(h) = 0$ as $\boldsymbol{\alpha}_j$ does not depend on $\rv_k(h)$. 
We thus consider the case where $k = j$. 
When $g = h$, we have that
\begin{equation*}
\begin{split}
\frac{\partial \boldsymbol{\alpha}_j (h)}{\partial \rv_j(h)}
&= \frac{1}{\tau^2} \frac{ e^{\frac{ \rv_j(h) }{\tau^2}} }{\sum_{k \in \mathbb{F}_q} e^{\frac{ \rv_j(k) }{\tau^2}}}
- \frac{1}{\tau^2} \frac{ e^{\frac{ \rv_j(h) }{\tau^2}} e^{\frac{ \rv_j(h) }{\tau^2}} }{\left( \sum_{k \in \mathbb{F}_q} e^{\frac{ \rv_j(k) }{\tau^2}} \right)^2} \\
&= \frac{1}{\tau^2} \boldsymbol{\alpha}_j (h)
\left( 1 - \boldsymbol{\alpha}_j (h) \right) .
\end{split}
\end{equation*}
When $g \neq h$, we get
\begin{equation*}
\begin{split}
\frac{\partial \boldsymbol{\alpha}_j (g)}{\partial \rv_j(h)}
&= - \frac{1}{\tau^2} \frac{ e^{\frac{ \rv_j(g) }{\tau^2}} e^{\frac{ \rv_j(h) }{\tau^2}} }{\left( \sum_{\kappa \in \mathbb{F}_q} e^{\frac{ \rv_j(\kappa) }{\tau^2}} \right)^2}
= - \frac{1}{\tau^2} \boldsymbol{\alpha}_j (g) \boldsymbol{\alpha}_j (h) .
\end{split}
\end{equation*}
Collecting these findings and condensing them into a more compact form, we arrive at \eqref{equation:DerivativeAlpha}, which is the desired expression.
\end{IEEEproof}

% The following corollary immediately follows from Proposition~\ref{proposition:DerivativeAlphaEll}. 
\begin{corollary} \label{corollary:bounded_alpha_ell_derivatives}
The absolute value of the partial derivatives of $\boldsymbol{\alpha}_{j}$ with respect to $\rv_{k}(h)$ are bounded by
\begin{equation}
    \left|\frac{\partial\boldsymbol{\alpha}_{j}}{\partial\rv_{k}(h)}\right| \leq \frac{1}{4\tau^2}
\end{equation}
where $\tau > 0$ is the standard deviation of the effective observation. 
\end{corollary}

The proof of this corollary is trivial when $\boldsymbol{\alpha}_{j}$ is a valid probability vector, as is the case in this article. 
% Note that the absolute value of every partial derivative in Lemma \ref{lemma:DerivativeAlphaEll} is bounded by $1/(4 \tau^2)$ because $\boldsymbol{\alpha}_j$ is a probability vector.
% Also note that the partial derivatives of $\boldsymbol{\alpha}_j$ with respect to $\rv_k(h)$ are zero for all $k \neq j$ and all $h \in \mathbb{F}_q$.
We also note that, based on the state evolution of AMP, $\tau^2 \geq \sigma^2$ at every iteration irrespective of the iteration number.
We can therefore establish a uniform bound across iterations.
We are now ready to show that the absolute value of \eqref{equation:JacobianMatrixEntries} is bounded whenever $\ell = j$, i.e., at the root level of the computation tree.

\begin{proposition}[Root Derivatives] \label{lemma:DerivativeEllBound}
The partial derivatives of $\hat{\sv}_{\ell} \left(\rv, g \right)$ with respect to $\rv_{\ell} (h)$ are given by
\begin{equation} \label{equation:DerivativeSell}
\frac{\partial \hat{\sv}_{\ell} \left( \rv \right)}{\partial \rv_{\ell} (h)}
= \frac{1}{\tau^2} \hat{\sv}_{\ell} \left( \rv, h \right) \left( \ev_h - \hat{\sv}_{\ell} \left( \rv \right) \right)
\qquad \forall h \in \mathbb{F}_{q}
\end{equation}
where $\tau > 0$ is the standard deviation of the effective observation.
\end{proposition}
\begin{IEEEproof}
Leveraging Proposition~\ref{proposition:DerivativeAlphaEll} and denoting the standard inner product by $\langle\cdot, \cdot\rangle$, we have
\begin{equation*}
\begin{split}
&\frac{\partial \hat{\sv}_{\ell} \left( \rv \right)}{\partial \rv_{\ell} (h)}
= \frac{\partial}{\partial \rv_{\ell} (h)} \frac{ \boldsymbol{\alpha}_{\ell} \circ \operatorname*{\bigcirc}_{p \in N(v_{\ell})} \boldsymbol{\mu}_{c_p \to v_{\ell}} }
{ \left\| \boldsymbol{\alpha}_{\ell} \circ \operatorname*{\bigcirc}_{p \in N(v_{\ell})} \boldsymbol{\mu}_{c_p \to v_{\ell}} \right\|_1 } \\
% &= \frac{ \frac{\partial \bar{\boldsymbol{\alpha}}_{\ell}}{\partial \rv_{\ell} (h)}
% \circ \left( \operatorname*{\bigcirc}_{c_p \in \mathrm{N}(v_{\ell})} \boldsymbol{\mu}_{c_p \to v_{\ell}} \right) }
% { \left\| \bar{\boldsymbol{\alpha}}_{\ell} \circ \left( \operatorname*{\bigcirc}_{c_p \in \mathrm{N}(v_{\ell})} \boldsymbol{\mu}_{c_p \to v_{\ell}} \right) \right\|_1 } \\
% &- \hat{\sv}_{\ell} \left( \rv \right) \frac{ \left\langle \frac{\partial \bar{\boldsymbol{\alpha}}_{\ell}}{\partial \rv_{\ell} (h)},
% \operatorname*{\bigcirc}_{c_p \in \mathrm{N}(v_{\ell})} \boldsymbol{\mu}_{c_p \to v_{\ell}} \right\rangle }
% { \left\| \bar{\boldsymbol{\alpha}}_{\ell} \circ \left( \operatorname*{\bigcirc}_{c_p \in \mathrm{N}(v_{\ell})} \boldsymbol{\mu}_{c_p \to v_{\ell}} \right) \right\|_1 } \\
&= \frac{\boldsymbol{\alpha}_{\ell} (h)}{\tau^2} \frac{ \left( \ev_h - \boldsymbol{\alpha}_{\ell} \right)
\circ \left( \operatorname*{\bigcirc}_{c_p \in \mathrm{N}(v_{\ell})} \boldsymbol{\mu}_{c_p \to v_{\ell}} \right) }
{ \left\| \boldsymbol{\alpha}_{\ell} \circ \left( \operatorname*{\bigcirc}_{c_p \in \mathrm{N}(v_{\ell})} \boldsymbol{\mu}_{c_p \to v_{\ell}} \right) \right\|_1 } \\
&- \frac{\boldsymbol{\alpha}_{\ell} (h)}{\tau^2} \hat{\sv}_{\ell} \left( \rv \right) \frac{ \left\langle \ev_h - \boldsymbol{\alpha}_{\ell},
\operatorname*{\bigcirc}_{c_p \in \mathrm{N}(v_{\ell})} \boldsymbol{\mu}_{c_p \to v_{\ell}} \right\rangle }
{ \left\| \boldsymbol{\alpha}_{\ell} \circ \left( \operatorname*{\bigcirc}_{c_p \in \mathrm{N}(v_{\ell})} \boldsymbol{\mu}_{c_p \to v_{\ell}} \right) \right\|_1 } \\
&= \frac{\boldsymbol{\alpha}_{\ell} (h)}{\tau^2}
\frac{ \ev_h \circ \left( \operatorname*{\bigcirc}_{c_p \in \mathrm{N}(v_{\ell})} \boldsymbol{\mu}_{c_p \to v_{\ell}} \right) }
{ \left\| \boldsymbol{\alpha}_{\ell} \circ \left( \operatorname*{\bigcirc}_{c_p \in \mathrm{N}(v_{\ell})} \boldsymbol{\mu}_{c_p \to v_{\ell}} \right) \right\|_1 } \\
&- \frac{\boldsymbol{\alpha}_{\ell} (h)}{\tau^2} \hat{\sv}_{\ell} \left( \rv \right)
\frac{ \left\langle \ev_h, \operatorname*{\bigcirc}_{c_p \in \mathrm{N}(v_{\ell})} \boldsymbol{\mu}_{c_p \to v_{\ell}} \right\rangle }
{ \left\| \boldsymbol{\alpha}_{\ell} \circ \left( \operatorname*{\bigcirc}_{c_p \in \mathrm{N}(v_{\ell})} \boldsymbol{\mu}_{c_p \to v_{\ell}} \right) \right\|_1 } \\
% &= \frac{1}{\tau^2} \frac{ \bar{\boldsymbol{\alpha}}_{\ell} (h) \prod_{c_p \in \mathrm{N}(v_{\ell})} \boldsymbol{\mu}_{c_p \to v_{\ell}} (h) }
% { \left\| \bar{\boldsymbol{\alpha}}_{\ell} \circ \left( \operatorname*{\bigcirc}_{c_p \in \mathrm{N}(v_{\ell})} \boldsymbol{\mu}_{c_p \to v_{\ell}} \right) \right\|_1 } \ev_h \\
% &- \frac{1}{\tau^2} \frac{ \boldsymbol{\alpha}_{\ell} (h) \prod_{c_p \in \mathrm{N}(v_{\ell})} \boldsymbol{\mu}_{c_p \to v_{\ell}} (h)}
% { \left\| \boldsymbol{\alpha}_{\ell} \circ \left( \operatorname*{\bigcirc}_{c_p \in \mathrm{N}(v_{\ell})} \boldsymbol{\mu}_{c_p \to v_{\ell}} \right) \right\|_1 } \hat{\sv}_{\ell} \left( \rv \right) \\
% &= \frac{1}{\tau^2} \hat{\sv}_{\ell} \left( \rv, h \right) \ev_h
% - \frac{1}{\tau^2} \hat{\sv}_{\ell} \left( \rv, h \right) \hat{\sv}_{\ell} \left( \rv \right) \\
&= \frac{1}{\tau^2} \hat{\sv}_{\ell} \left( \rv, h \right) \left( \ev_h - \hat{\sv}_{\ell} \left( \rv \right) \right),
\end{split}
\end{equation*}
which is the desired expression. 
\end{IEEEproof}
\begin{corollary} \label{corollary:bounded_root_derivatives}
The absolute value of the partial derivatives of $\hat{\sv}_{\ell}\left(\rv\right)$ with respect to $\rv_{\ell}(h)$ are bounded by
\begin{equation}
    \left|\frac{\partial\hat{\sv}_{\ell}\left(\rv\right)}{\partial\rv_{\ell}(h)}\right| \leq \frac{1}{4\tau^2}
\end{equation}
\end{corollary}
The proof of this corollary follows that of Corollary \ref{corollary:bounded_alpha_ell_derivatives} because like $\boldsymbol{\alpha}_{j}$, $\hat{\sv}_{\ell}\left(\rv\right)$ forms a valid probability vector. 

Proposition~\ref{lemma:DerivativeEllBound} offers a blueprint for the general result we wish to establish.
Yet, the situation gets more complicated when $\ell \neq j$ because we have to involve the message passing rules.
In doing so, we obtain a key intermediate result using mathematical induction.
We start with the variable node closest to the root node, and then progress outward step by step.

To circumvent a notational nightmare, we restrict the proof to cases where all edge weights are equal to $1 \in \mathbb{F}_{q}$.
Conceptually, the edge can be interpreted as permutations on the belief vectors.
From the point of view of bounding partial derivatives, this is a benign operation, yet the accounting that comes with permutations is dreadful, hence our focus on the simpler case.
Moving forward, we assume the following condition.

\begin{condition} \label{condition:Edge1}
All edge weights within the factor graph of the LDPC outer code are equal to $1 \in \mathbb{F}_{q}$.
\end{condition}

The extension of the following propositions to the case with arbitrary edge weights (i.e., beyond Condition~\ref{condition:Edge1}) is conceptually straightforward.

\begin{proposition} \label{proposition:DerivativeJGammaBound}
Suppose Condition~\ref{condition:Sub-Girth AMP-BP} holds and let $v_j$ be a descendant of root node $v_{\ell}$ in the computation tree.
Moreover, let $c_p \in N(v_{\ell})$ be the unique check neighbor of $v_{\ell}$ on the path from $v_{\ell}$ to $v_j$ within the tree.
Then, there exists vector $\boldsymbol{\nu}$, with $\zerov \preceq \boldsymbol{\nu} \preceq \boldsymbol{\mu}_{c_p \to v_{\ell}}$, such that the partial derivative of $\boldsymbol{\mu}_{c_p \to v_{\ell}}$ with respect to $\rv_j (h)$ is given by
\begin{equation} \label{equation:DerivativeMuJ}
\frac{\partial \boldsymbol{\mu}_{c_p \to v_{\ell}}}{\partial \rv_j (h)}
= \frac{1}{\tau^2} \left( \boldsymbol{\nu} - \left\| \boldsymbol{\nu} \right\|_1
\boldsymbol{\mu}_{c_p \to v_{\ell}} \right),
\end{equation}
where $\tau > 0$ is the standard deviation of the effective observation. 
Here, $\preceq$ denotes elementwise comparison of the entries in the vector. 
\end{proposition}
\begin{IEEEproof}
Under condition~\ref{condition:Sub-Girth AMP-BP}, we know that $v_j$ appears at most once within the computation tree rooted at $v_{\ell}$. 
Thus, we establish \eqref{equation:DerivativeMuJ} via mathematical induction on the distance between $v_{\ell}$ and its descendant $v_j$ on the computation tree.
The distance that we are interested in only considers the number of variable nodes between $v_{\ell}$ and $v_j$. 
Before beginning, we point out that if $v_j$ is not a descendant of $v_{\ell}$, then the corresponding partial derivatives vanish and the claim is immediate.

We begin with generic results that are useful for both the base case and the inductive step.
Let $v_j$ be a descendant of $v_{\ell}$ and let $v_k$ be the variable child of $v_{\ell}$ on the path from $v_{\ell}$ to $v_j$.
Let $c_p$ be the unique check node in $N(v_{\ell}) \cap N(v_k)$ and let $c_{\varrho}$ be the unique check node child of $v_k$ on the path from $v_k$ to $v_j$; if $v_k = v_j$, let $\varrho = 0$. 
Then, we have that
% Consider the partial derivatives of probability vector $\boldsymbol{\mu}_{v_k \to c_p}$.
% Based on the operations that take place at the variable nodes, we can write
\begin{equation} \label{equation:DerivativeMukp}
\begin{split}
&\frac{\partial \boldsymbol{\mu}_{v_k \to c_p}}{\partial \rv_j (h)}
= \frac{\partial}{\partial \rv_j (h)}
\frac{ \operatorname*{\bigcirc}_{c_{\xi} \in N_0(v_k) \setminus c_{p}} \boldsymbol{\mu}_{c_{\xi} \to v_k} }
{ \left\| \operatorname*{\bigcirc}_{c_{\xi} \in N_0(v_k) \setminus c_{p}}  \boldsymbol{\mu}_{c_{\xi} \to v_k} \right\|_1 } .
% &= \frac{ \frac{\partial \boldsymbol{\mu}_{c_{\rho} \to v_k}}{\partial \rv_j (h)} \circ
% \left( \operatorname*{\bigcirc}_{c_{\varrho} \in N_0(v_k) \setminus c_{p}, c_{\rho}} \boldsymbol{\mu}_{c_{\varrho} \to v_k} \right) }
% { \left\langle \boldsymbol{\mu}_{c_{\rho} \to v_k}, \operatorname*{\bigcirc}_{c_{\varrho} \in N_0(v_k) \setminus c_{p}, c_{\rho}} \boldsymbol{\mu}_{c_{\varrho} \to v_k} \right\rangle } \\
% &- \boldsymbol{\mu}_{v_k \to c_p} \frac{ \left\langle \frac{\partial \boldsymbol{\mu}_{c_{\rho} \to v_k}}{\partial \rv_j (h)},
% \operatorname*{\bigcirc}_{c_{\varrho} \in N_0(v_k) \setminus c_{p}, c_{\rho}} \boldsymbol{\mu}_{c_{\varrho} \to v_k} \right\rangle }
% { \left\langle \boldsymbol{\mu}_{c_{\rho} \to v_k}, \operatorname*{\bigcirc}_{c_{\varrho} \in N_0(v_k) \setminus c_{p}, c_{\rho}} \boldsymbol{\mu}_{c_{\varrho} \to v_k} \right\rangle } .
\end{split}
\end{equation}
We can also examine the partial derivatives of the probability vector $\boldsymbol{\mu}_{c_{\varrho} \to v_k}$.
Suppose $v_k \neq v_j$ and let $v_o$ be the unique variable child of $v_k$ on the path from $v_{\ell}$ to $v_j$.
Using the $\mathbb{F}_q$ convolution, we have
\begin{equation} \label{equation:DerivativeMupl}
\begin{split}
&\frac{\partial \boldsymbol{\mu}_{c_{\varrho} \to v_k}}{\partial \rv_j (h)}
= \frac{\partial}{\partial \rv_j (h)} \left( \bigodot_{v_l \in N(c_{\varrho}) \setminus v_k} \mkern-18mu \boldsymbol{\mu}_{v_l \to c_{\varrho}} \right) \\
&= \frac{\partial \boldsymbol{\mu}_{v_o \to c_{\varrho}}}{\partial \rv_j (h)} \odot
\left( \bigodot_{v_l \in N(c_{\varrho}) \setminus \{ v_k, v_o \}} \mkern-36mu \boldsymbol{\mu}_{v_l \to c_{\varrho}} \right) \\
&= \frac{\partial \boldsymbol{\mu}_{v_o \to c_{\varrho}}}{\partial \rv_j (h)} \odot
\boldsymbol{\nu}_{c_{\varrho} \setminus v_k, v_o} .
\end{split}
\end{equation}
We emphasize that $\boldsymbol{\nu}_{c_{\varrho} \setminus v_k, v_o}$, as defined implicitly above, is a probability distribution.

Having established these results, we turn our attention to the base case where $v_j$ is a variable child of $v_{\ell}$ ($v_k = v_j$, $\varrho = 0$). 
% , and $\boldsymbol{\mu}_{c_{\varrho} \to v_k} = \boldsymbol{\mu}_{c_0 \to v_j} = \boldsymbol{\alpha}_j$.
Applying \eqref{equation:DerivativeMukp} and Proposition~\ref{proposition:DerivativeAlphaEll}, we obtain
\begin{equation}
\begin{split}
&\frac{\partial \boldsymbol{\mu}_{v_j \to c_p}}{\partial \rv_j (h)}
= \frac{\partial}{\partial \rv_j (h)}
\frac{ \boldsymbol{\alpha}_{j}\circ\left(\operatorname*{\bigcirc}_{c_{\xi} \in N(v_j) \setminus c_{p}} \boldsymbol{\mu}_{c_{\xi} \to v_j} \right)}
{ \left\| \boldsymbol{\alpha}_{j}\circ\left(\operatorname*{\bigcirc}_{c_{\xi} \in N(v_j) \setminus c_{p}}  \boldsymbol{\mu}_{c_{\xi} \to v_j}\right) \right\|_1 } \\
% &= \frac{ \frac{\partial \boldsymbol{\mu}_{c_0 \to v_j}}{\partial \rv_j (h)} \circ
% \left( \operatorname*{\bigcirc}_{c_{\varrho} \in N_0(v_j) \setminus c_p, c_0} \boldsymbol{\mu}_{c_{\varrho} \to v_j} \right) }
% { \left\langle \boldsymbol{\mu}_{c_0 \to v_j}, \operatorname*{\bigcirc}_{c_{\varrho} \in N_0(v_j) \setminus c_p, c_0} \boldsymbol{\mu}_{c_{\varrho} \to v_j} \right\rangle } \\
% &- \boldsymbol{\mu}_{v_j \to c_p} \frac{ \left\langle \frac{\partial \boldsymbol{\mu}_{c_0 \to v_j}}{\partial \rv_j (h)},
% \operatorname*{\bigcirc}_{c_{\varrho} \in N_0(v_j) \setminus c_p, c_0} \boldsymbol{\mu}_{c_{\varrho} \to v_j} \right\rangle }
% { \left\langle \boldsymbol{\mu}_{c_0 \to v_j}, \operatorname*{\bigcirc}_{c_{\varrho} \in N_0(v_j) \setminus c_p, c_0} \boldsymbol{\mu}_{c_{\varrho} \to v_j} \right\rangle } \\
&= \frac{ \frac{\partial \boldsymbol{\alpha}_j}{\partial \rv_j (h)} \circ
\left( \operatorname*{\bigcirc}_{c_{\xi} \in N(v_j) \setminus c_p} \boldsymbol{\mu}_{c_{\xi} \to v_j} \right) }
{ \left\langle \boldsymbol{\alpha}_j, \operatorname*{\bigcirc}_{c_{\xi} \in N(v_j) \setminus c_p} \boldsymbol{\mu}_{c_{\xi} \to v_j} \right\rangle } \\
&- \boldsymbol{\mu}_{v_j \to c_p} \frac{ \left\langle \frac{\partial \boldsymbol{\alpha}_j}{\partial \rv_j (h)},
\operatorname*{\bigcirc}_{c_{\xi} \in N(v_j) \setminus c_p} \boldsymbol{\mu}_{c_{\xi} \to v_j} \right\rangle }
{ \left\langle \boldsymbol{\alpha}_j, \operatorname*{\bigcirc}_{c_{\xi} \in N(v_j) \setminus c_p} \boldsymbol{\mu}_{c_{\xi} \to v_j} \right\rangle } \\
% &= \frac{\boldsymbol{\alpha}_j (h)}{\tau^2} \frac{ \left( \ev_h - \boldsymbol{\alpha}_j \right) \circ
% \left( \operatorname*{\bigcirc}_{c_{\varrho} \in N(v_j) \setminus c_p, c_0} \boldsymbol{\mu}_{c_{\varrho} \to v_j} \right) }
% { \left\langle \boldsymbol{\alpha}_j, \operatorname*{\bigcirc}_{c_{\varrho} \in N(v_j) \setminus c_p, c_0} \boldsymbol{\mu}_{c_{\varrho} \to v_j} \right\rangle } \\
% &- \frac{\boldsymbol{\alpha}_j (h)}{\tau^2} \boldsymbol{\mu}_{v_j \to c_p}
% \frac{ \left\langle \ev_h - \boldsymbol{\alpha}_j ,
% \operatorname*{\bigcirc}_{c_{\varrho} \in N(v_j) \setminus c_p, c_0} \boldsymbol{\mu}_{c_{\varrho} \to v_j} \right\rangle }
% { \left\langle \boldsymbol{\alpha}_j, \operatorname*{\bigcirc}_{c_{\varrho} \in N(v_j) \setminus c_p, c_0} \boldsymbol{\mu}_{c_{\varrho} \to v_j} \right\rangle } \\
&= \frac{1}{\tau^2} \frac{ \boldsymbol{\alpha}_j (h) \ev_h \circ
\left( \operatorname*{\bigcirc}_{c_{\xi} \in N(v_j) \setminus c_p} \boldsymbol{\mu}_{c_{\xi} \to v_j} \right) }
{ \left\langle \boldsymbol{\alpha}_j, \operatorname*{\bigcirc}_{c_{\xi} \in N(v_j) \setminus c_p} \boldsymbol{\mu}_{c_{\xi} \to v_j} \right\rangle } \\
&- \frac{1}{\tau^2} \boldsymbol{\mu}_{v_j \to c_p}
\frac{ \left\langle \boldsymbol{\alpha}_j (h) \ev_h ,
\operatorname*{\bigcirc}_{c_{\xi} \in N(v_j) \setminus c_p} \boldsymbol{\mu}_{c_{\xi} \to v_j} \right\rangle }
{ \left\langle \boldsymbol{\alpha}_j, \operatorname*{\bigcirc}_{c_{\xi} \in N(v_j) \setminus c_p} \boldsymbol{\mu}_{c_{\xi} \to v_j} \right\rangle } \\
&= \frac{1}{\tau^2} \left( \boldsymbol{\nu}_{v_j \to c_p}
- \left\| \boldsymbol{\nu}_{v_j \to c_p} \right\|_1 \boldsymbol{\mu}_{v_j \to c_p} \right),
\end{split}
\end{equation}
% In this derivation, we have employed the fact that
% \begin{equation*}
% \boldsymbol{\mu}_{v_j \to c_p} = \frac{ \bar{\boldsymbol{\alpha}}_j \circ
% \left( \operatorname*{\bigcirc}_{c_{\varrho} \in N_0(v_j) \setminus c_p, c_0} \boldsymbol{\mu}_{c_{\varrho} \to v_j} \right) }
% { \left\langle \bar{\boldsymbol{\alpha}}_j, \operatorname*{\bigcirc}_{c_{\varrho} \in N_0(v_j) \setminus c_p, c_0} \boldsymbol{\mu}_{c_{\varrho} \to v_j} \right\rangle }
% \end{equation*}
% to cancel out two terms.
where
\begin{equation*}
\boldsymbol{\nu}_{v_j \to c_p} = \frac{ \boldsymbol{\alpha}_j (h) \ev_h \circ
\left( \operatorname*{\bigcirc}_{c_{\xi} \in N(v_j) \setminus c_p} \boldsymbol{\mu}_{c_{\xi} \to v_j} \right) }
{ \left\langle \boldsymbol{\alpha}_j, \operatorname*{\bigcirc}_{c_{\xi} \in N(v_j) \setminus c_p} \boldsymbol{\mu}_{c_{\xi} \to v_j} \right\rangle } .
\end{equation*}
By construction, we have $\zerov \preceq \boldsymbol{\nu}_{v_j \to c_p} \preceq \boldsymbol{\mu}_{v_j \to c_p}$.
% This can be seen by comparing the two equations above.
% 
We turn to the second graph operation and apply \eqref{equation:DerivativeMupl}, which yields
\begin{equation} \label{equation:DerivativeMu0jh}
\begin{split}
&\frac{\partial \boldsymbol{\mu}_{c_p \to v_{\ell}}}{\partial \rv_j(h)}
= \frac{\partial \boldsymbol{\mu}_{v_j \to c_p}}{\partial \rv_j (h)}
\odot \boldsymbol{\nu}_{c_p \setminus v_{\ell}, v_j} \\
&= \frac{1}{\tau^2} \left( \boldsymbol{\nu}_{v_j \to c_p} - \left\| \boldsymbol{\nu}_{v_j \to c_p} \right\|_1 \boldsymbol{\mu}_{v_j \to c_p} \right) 
\odot \boldsymbol{\nu}_{c_p \setminus v_{\ell}, v_j} \\
&= \frac{1}{\tau^2} \left( \boldsymbol{\nu}_{v_j \to c_p} \odot \boldsymbol{\nu}_{c_p \setminus v_{\ell}, v_j}
- \left\| \boldsymbol{\nu}_{v_j \to c_p} \right\|_1 \boldsymbol{\mu}_{c_p \to v_{\ell}} \right) .
\end{split}
\end{equation}
Thus, in this case, we take $\boldsymbol{\nu} = \boldsymbol{\nu}_{v_j \to c_p} \odot \boldsymbol{\nu}_{c_p \setminus v_{\ell}, v_j}$ as a suitable vector.
Based on the fact that $\boldsymbol{\nu}_{c_p \setminus v_{\ell}, v_j}$ is a probability vector, together with the aforementioned component-wise ordering, we gather that
\begin{equation*}
\begin{split}
\zerov &\preceq \boldsymbol{\nu} = \boldsymbol{\nu}_{v_j \to c_p} \odot \boldsymbol{\nu}_{c_p \setminus v_{\ell}, v_j} \\
&\preceq \boldsymbol{\mu}_{v_j \to c_p} \odot \boldsymbol{\nu}_{c_p \setminus v_{\ell}, v_j}
= \boldsymbol{\mu}_{c_p \to v_{\ell}} .
\end{split}
\end{equation*}
Moreover, leveraging the properties of the $\mathbb{F}_q$ convolution for non-negative vectors, we get
\begin{equation*}
\left\| \boldsymbol{\nu} \right\|_1
= \left\| \boldsymbol{\nu}_{v_j \to c_p} \right\|_1 \left\| \boldsymbol{\nu}_{c_p \setminus v_{\ell}, v_j} \right\|_1
= \left\| \boldsymbol{\nu}_{v_j \to c_p} \right\|_1 .
\end{equation*}
Thus, for this choice of $\boldsymbol{\nu}$, we arrive at
\begin{equation} \label{equation:DerivativeMu0jhNu}
\frac{\partial \boldsymbol{\mu}_{c_p \to v_{\ell}}}{\partial \rv_j(h)}
= \frac{1}{\tau^2} \left( \boldsymbol{\nu} - \left\| \boldsymbol{\nu} \right\|_1
\boldsymbol{\mu}_{c_p \to v_{\ell}} \right) ,
\end{equation}
as claimed.
That is, the base case conforms to the structure of Proposition~\ref{proposition:DerivativeJGammaBound}.

We now consider the inductive step in our proof. 
As our hypothesis, we assume that \eqref{equation:DerivativeMuJ} holds for all computation trees wherein the distance between the root node and $v_j$ is less than or equal to $\gamma \in \mathbb{N}$.
Consider a rooted computation tree and suppose the distance between $v_{\ell}$ and its descendant $v_j$ in the tree is exactly $\gamma + 1$.
Under Condition~\ref{condition:Sub-Girth AMP-BP}, there is a unique path from $v_{\ell}$ to node $v_j$.
Let $v_k$ be the variable child of $v_{\ell}$ that is also an ascendant of $v_j$, and denote the unique check node that connects the two by $c_p \in N(v_{\ell}) \cap N(v_k)$.
Furthermore, let $c_{\varrho} \in N(v_k)$ be the unique check node within this neighborhood that is an ascendant of $v_j$ on the computation tree.
Finally, let $v_o \in N(c_{\varrho})$ be the unique variable child of $v_k$ that is also an ascendant of $v_j$ (or, perhaps, $v_j$ itself).

The sub-tree starting at $v_k$ can be viewed as a rooted tree containing $v_j$; the graph distance between these two variable nodes within the sub-tree is exactly $\gamma$.
As such, our inductive hypothesis applies.
That is, there exists vector $\boldsymbol{\nu}$ such that $\zerov \preceq \boldsymbol{\nu} \preceq \boldsymbol{\mu}_{c_{\varrho} \to v_k}$ where the partial derivative of $\boldsymbol{\mu}_{c_{\varrho} \to v_k}$ with respect to $\rv_j (h)$ is equal to
\begin{equation}
\frac{\partial \boldsymbol{\mu}_{c_{\varrho} \to v_k}}{\partial \rv_j (h)}
= \frac{1}{\tau^2} \left( \boldsymbol{\nu} - \left\| \boldsymbol{\nu} \right\|_1
\boldsymbol{\mu}_{c_{\varrho} \to v_k} \right) .
\end{equation}

Applying \eqref{equation:DerivativeMukp} and our inductive hypothesis, we have that
\begin{equation} \label{equation:paritalMukp}
\begin{split}
&\frac{\partial \boldsymbol{\mu}_{v_k \to c_p}}{\partial \rv_j (h)}
= \frac{\partial}{\partial \rv_j (h)}
\frac{\operatorname*{\bigcirc}_{c_{\xi} \in N_0(v_k) \setminus c_{p}} \boldsymbol{\mu}_{c_{\xi} \to v_k} }
{ \left\| \operatorname*{\bigcirc}_{c_{\xi} \in N_0(v_k) \setminus c_{p}}  \boldsymbol{\mu}_{c_{\xi} \to v_k} \right\|_1 } \\
&= \frac{ \frac{\partial \boldsymbol{\mu}_{c_{\varrho} \to v_k}}{\partial \rv_j (h)} \circ
\left( \operatorname*{\bigcirc}_{c_{\xi} \in N_0(v_k) \setminus c_p, c_{\varrho}} \boldsymbol{\mu}_{c_{\xi} \to v_k} \right) }
{ \left\langle \boldsymbol{\mu}_{c_{\varrho} \to v_k}, \operatorname*{\bigcirc}_{c_{\xi} \in N_0(v_k) \setminus c_p, c_{\varrho}} \boldsymbol{\mu}_{c_{\xi} \to v_k} \right\rangle } \\
&- \boldsymbol{\mu}_{v_k \to c_p} \frac{ \left\langle \frac{\partial \boldsymbol{\mu}_{c_{\varrho} \to v_k}}{\partial \rv_j (h)},
\operatorname*{\bigcirc}_{c_{\xi} \in N_0(v_k) \setminus c_p, c_{\varrho}} \boldsymbol{\mu}_{c_{\xi} \to v_k} \right\rangle }
{ \left\langle \boldsymbol{\mu}_{c_{\varrho} \to v_k}, \operatorname*{\bigcirc}_{c_{\xi} \in N_0(v_k) \setminus c_p, c_{\varrho}} \boldsymbol{\mu}_{c_{\xi} \to v_k} \right\rangle } \\
% &= \frac{1}{\tau^2} \frac{ \left( \boldsymbol{\nu} - \left\| \boldsymbol{\nu} \right\|_1 \boldsymbol{\mu}_{c_{\rho} \to v_k} \right) \circ
% \left( \operatorname*{\bigcirc}_{c_{\varrho} \in N_0(v_k) \setminus c_p, c_{\rho}} \boldsymbol{\mu}_{c_{\varrho} \to v_k} \right) }
% { \left\langle \boldsymbol{\mu}_{c_{\rho} \to v_k}, \operatorname*{\bigcirc}_{c_{\varrho} \in N_0(v_k) \setminus c_p, c_{\rho}} \boldsymbol{\mu}_{c_{\varrho} \to v_k} \right\rangle } \\
% &- \frac{1}{\tau^2} \boldsymbol{\mu}_{v_k \to c_p}
% \frac{ \left\langle \boldsymbol{\nu} - \left\| \boldsymbol{\nu} \right\|_1 \boldsymbol{\mu}_{c_{\rho} \to v_k},
% \operatorname*{\bigcirc}_{c_{\varrho} \in N_0(v_k) \setminus c_p, c_{\rho}} \boldsymbol{\mu}_{c_{\varrho} \to v_k} \right\rangle }
% { \left\langle \boldsymbol{\mu}_{c_{\rho} \to v_k}, \operatorname*{\bigcirc}_{c_{\varrho} \in N_0(v_k) \setminus c_p, c_{\rho}} \boldsymbol{\mu}_{c_{\varrho} \to v_k} \right\rangle } \\
&= \frac{1}{\tau^2} \frac{ \boldsymbol{\nu} \circ
\left( \operatorname*{\bigcirc}_{c_{\xi} \in N_0(v_k) \setminus c_p, c_{\varrho}} \boldsymbol{\mu}_{c_{\xi} \to v_k} \right) }
{ \left\langle \boldsymbol{\mu}_{c_{\varrho} \to v_k}, \operatorname*{\bigcirc}_{c_{\xi} \in N_0(v_k) \setminus c_p, c_{\varrho}} \boldsymbol{\mu}_{c_{\xi} \to v_k} \right\rangle } \\
&- \frac{1}{\tau^2} \boldsymbol{\mu}_{v_k \to c_p}
\frac{ \left\langle \boldsymbol{\nu},
\operatorname*{\bigcirc}_{c_{\xi} \in N_0(v_k) \setminus c_p, c_{\varrho}} \boldsymbol{\mu}_{c_{\xi} \to v_k} \right\rangle }
{ \left\langle \boldsymbol{\mu}_{c_{\varrho} \to v_k}, \operatorname*{\bigcirc}_{c_{\xi} \in N_0(v_k) \setminus c_p, c_{\varrho}} \boldsymbol{\mu}_{c_{\xi} \to v_k} \right\rangle } \\
&= \frac{1}{\tau^2} \left( \boldsymbol{\nu}_{v_k \to c_p}
- \left\| \boldsymbol{\nu}_{v_k \to c_p} \right\|_1 \boldsymbol{\mu}_{v_k \to c_p} \right) ,
\end{split}
\end{equation}
where we have utilized the shorthand notation
\begin{equation*}
\boldsymbol{\nu}_{v_k \to c_p} = \frac{ \boldsymbol{\nu} \circ
\left( \operatorname*{\bigcirc}_{c_{\xi} \in N_0(v_k) \setminus c_p, c_{\varrho}} \boldsymbol{\mu}_{c_{\xi} \to v_k} \right) }
{ \left\langle \boldsymbol{\mu}_{c_{\varrho} \to v_k}, \operatorname*{\bigcirc}_{c_{\xi} \in N_0(v_k) \setminus c_p, c_{\varrho}} \boldsymbol{\mu}_{c_{\xi} \to v_k} \right\rangle } .
\end{equation*}
We emphasize that two of the terms in the derivation above cancel out, as before.
Furthermore, by construction, we immediately obtain $\zerov \preceq \boldsymbol{\nu}_{v_k \to c_p} \preceq \boldsymbol{\mu}_{v_k \to c_p}$.
These observations closely parallel the description for the base case.

The derivation of the second graph operation for the inductive step is in complete analogy with the base case, except for labeling.
Specifically, we apply \eqref{equation:DerivativeMupl} and obtain
\begin{equation}
\begin{split}
&\frac{\partial \boldsymbol{\mu}_{c_p \to v_{\ell}}}{\partial \rv_j(h)}
= \frac{\partial \boldsymbol{\mu}_{v_k \to c_p}}{\partial \rv_j (h)}
\odot \boldsymbol{\nu}_{c_p \setminus v_{\ell}, v_k} \\
&= \frac{1}{\tau^2} \left( \boldsymbol{\nu}_{v_k \to c_p} - \left\| \boldsymbol{\nu}_{v_k \to c_p} \right\|_1 \boldsymbol{\mu}_{v_k \to c_p} \right) 
\odot \boldsymbol{\nu}_{c_p \setminus v_{\ell}, v_k} \\
&= \frac{1}{\tau^2} \left( \boldsymbol{\nu}_{v_k \to c_p} \odot \boldsymbol{\nu}_{c_p \setminus v_{\ell}, v_k}
- \left\| \boldsymbol{\nu}_{v_k \to c_p} \right\|_1 \boldsymbol{\mu}_{c_p \to v_{\ell}} \right) .
\end{split}
\end{equation}
For the inductive step, we define $\boldsymbol{\nu}' = \boldsymbol{\nu}_{v_k \to c_p} \odot \boldsymbol{\nu}_{c_p \setminus v_{\ell}, v_k}$ as the candidate vector.
Based on component-wise ordering, we can write
\begin{equation*}
\begin{split}
\zerov &\preceq \boldsymbol{\nu}' = \boldsymbol{\nu}_{v_k \to c_p} \odot \boldsymbol{\nu}_{c_p \setminus v_{\ell}, v_k} \\
&\preceq \boldsymbol{\mu}_{v_k \to c_p} \odot \boldsymbol{\nu}_{c_p \setminus v_{\ell}, v_k}
= \boldsymbol{\mu}_{c_p \to v_{\ell}} .
\end{split}
\end{equation*}
As before, we have that
\begin{equation*}
\left\| \boldsymbol{\nu}' \right\|_1
= \left\| \boldsymbol{\nu}_{v_k \to c_p} \right\|_1 \left\| \boldsymbol{\nu}_{c_p \setminus v_{\ell}, v_k} \right\|_1
= \left\| \boldsymbol{\nu}_{v_k \to c_p} \right\|_1 .
\end{equation*}
Hence, candidate vector $\boldsymbol{\nu}'$ is such that $\zerov \preceq \boldsymbol{\nu}' \preceq \boldsymbol{\mu}_{c_p \to v_{\ell}}$ and
\begin{equation}
\frac{\partial \boldsymbol{\mu}_{c_p \to v_{\ell}}}{\partial \rv_j(h)}
= \frac{1}{\tau^2} \left( \boldsymbol{\nu}' - \left\| \boldsymbol{\nu}' \right\|_1
\boldsymbol{\mu}_{c_p \to v_{\ell}} \right) .
\end{equation}
This completes the proof for Proposition~\ref{proposition:DerivativeJGammaBound}.
\end{IEEEproof}

We have nearly attained out goal of showing that the magnitudes of the entries in the Jacobian matrix of $\etav(\rv)$ with respect to $\rv$ are uniformly bounded.
To achieve the desired result, it sufficies to connect the partial derivative of the incoming message with the partial derivative of state estimate $\hat{\sv}_{\ell} \left( \rv, g \right)$.
This is accomplished below.

\begin{proposition} \label{proposition:DerivativeJBound}
Under Condition~\ref{condition:Sub-Girth AMP-BP}, the absolute value of the entries in the Jacobian are bounded by
\begin{equation} \label{equation:DerivativeJBound}
\left| \frac{\partial \hat{\sv}_{\ell} \left( \rv, g \right)}{\partial \rv_{j} (h)} \right|
\leq \frac{1}{\tau^2} \quad \forall g, h \in \mathbb{F}_q, \ell, j \in [L] ,
\end{equation}
where $\tau > 0$ is the standard deviation of the effective observation.
\end{proposition}
\begin{IEEEproof}
When $v_j$ does not appear in the rooted tree of $v_{\ell}$, the partial derivative is equal to zero and the result immediately follows. 
Furthermore, when $v_{\ell} = v_j$, the result follows from Corollary~\ref{corollary:bounded_root_derivatives}.
Thus, we can focus on the scenario wherein $v_j$ is a descendant of $v_{\ell}$.

Let $c_p$ be the unique check node in $N(v_{\ell})$ that lies on the path between $v_{\ell}$ and $v_j$.
By Proposition~\ref{proposition:DerivativeJGammaBound}, there exists vector $\boldsymbol{\nu}$, with $\zerov \preceq \boldsymbol{\nu} \preceq \boldsymbol{\mu}_{c_p \to v_{\ell}}$, such that the partial derivative of $\boldsymbol{\mu}_{c_p \to v_{\ell}}$ with respect to $\rv_j (h)$ is given by
\begin{equation}
\frac{\partial \boldsymbol{\mu}_{c_p \to v_{\ell}}}{\partial \rv_j (h)}
= \frac{1}{\tau^2} \left( \boldsymbol{\nu} - \left\| \boldsymbol{\nu} \right\|_1
\boldsymbol{\mu}_{c_p \to v_{\ell}} \right) .
\end{equation}
Drawing an analogy to \eqref{equation:paritalMukp}, we have
\begin{equation}
\begin{split}
&\frac{\partial \hat{\sv}_{\ell} \left( \rv \right)}{\partial \rv_j (h)}
= \frac{\partial}{\partial \rv_j (h)}
\frac{ \operatorname*{\bigcirc}_{c_{\xi} \in N_0(v_{\ell})} \boldsymbol{\mu}_{c_{\xi} \to v_{\ell}} }
{ \left\| \operatorname*{\bigcirc}_{c_{\xi} \in N_0(v_{\ell})}  \boldsymbol{\mu}_{c_{\xi} \to v_{\ell}} \right\|_1 } \\
&= \frac{1}{\tau^2} \frac{ \boldsymbol{\nu} \circ
\left( \operatorname*{\bigcirc}_{c_{\xi} \in N_0(v_{\ell}) \setminus c_p} \boldsymbol{\mu}_{c_{\xi} \to v_{\ell}} \right) }
{ \left\langle \boldsymbol{\mu}_{c_p \to v_{\ell}}, \operatorname*{\bigcirc}_{c_{\xi} \in N_0(v_{\ell}) \setminus c_p} \boldsymbol{\mu}_{c_{\xi} \to v_{\ell}} \right\rangle } \\
&- \frac{1}{\tau^2} \hat{\sv}_{\ell} \left( \rv \right)
\frac{ \left\langle \boldsymbol{\nu},
\operatorname*{\bigcirc}_{c_{\xi} \in N_0(v_{\ell}) \setminus c_p} \boldsymbol{\mu}_{c_{\xi} \to v_{\ell}} \right\rangle }
{ \left\langle \boldsymbol{\mu}_{c_p \to v_{\ell}}, \operatorname*{\bigcirc}_{c_{\xi} \in N_0(v_{\ell}) \setminus c_p} \boldsymbol{\mu}_{c_{\xi} \to v_{\ell}} \right\rangle } \\
&= \frac{1}{\tau^2} \left( \boldsymbol{\nu}_{v_{\ell}}
- \left\| \boldsymbol{\nu}_{v_{\ell}} \right\|_1 \hat{\sv}_{\ell} \left( \rv \right) \right) ,
\end{split}
\end{equation}
where we have implicitly defined
\begin{equation*}
\boldsymbol{\nu}_{v_{\ell}} = \frac{ \boldsymbol{\nu} \circ
\left( \operatorname*{\bigcirc}_{c_{\xi} \in N_0(v_{\ell}) \setminus c_p} \boldsymbol{\mu}_{c_{\xi} \to v_{\ell}} \right) }
{ \left\langle \boldsymbol{\mu}_{c_{p} \to v_{\ell}}, \operatorname*{\bigcirc}_{c_{\xi} \in N_0(v_{\ell}) \setminus c_p} \boldsymbol{\mu}_{c_{\xi} \to v_{\ell}} \right\rangle } .
\end{equation*}
% Again, two of the terms in the derivation cancel out.
We note that $\zerov \preceq \boldsymbol{\nu}_{v_{\ell}} \preceq \hat{\sv}_{\ell} \left( \rv \right)$.
% As a consequence, $\left\| \boldsymbol{\nu}_{v_{\ell}} \right\|_1 \leq 1$ and
% \begin{equation}
% % \begin{split}
% \frac{\partial \hat{\sv}_{\ell} \left( \rv \right)}{\partial \rv_j (h)}
% \geq - \frac{1}{\tau^2} \boldsymbol{\mu}_{v_{\ell} \to c_p} .
% \end{equation}
% Similarly, disregarding the negative terms, we have
Thus, we have that:
\begin{equation} \label{eq:thm_positive}
\frac{\partial \hat{\sv}_{\ell} \left( \rv \right)}{\partial \rv_j (h)}
\leq \frac{1}{\tau^2} \boldsymbol{\nu}_{v_{\ell}}
\leq \frac{1}{\tau^2} \hat{\sv}_{\ell} \left( \rv \right) .
\end{equation}
Since we are interested in bounding the absolute value of the partial derivatives, we also consider a lower bound. 
\begin{equation}
\frac{\partial \hat{\sv}_{\ell} \left( \rv \right)}{\partial \rv_j (h)}
\geq - \frac{1}{\tau^2} \left\| \boldsymbol{\nu}_{v_{\ell}} \right\|_1 \hat{\sv}_{\ell} \left( \rv \right)
\geq - \frac{1}{\tau^2} \hat{\sv}_{\ell} \left( \rv \right) .
\end{equation}

Combining these two observations with the properties of probability vectors, we obtain the desired expression.
\end{IEEEproof}

\begin{theorem}
\label{theorem:denoiser_lipschitz_continuous}
Under Condition~\ref{condition:Sub-Girth AMP-BP}, the BP-N denoiser presented in definition~\ref{definition:BPDenoiser} is Lipschitz continuous.
\end{theorem}
\begin{IEEEproof}
By Proposition~\ref{proposition:DerivativeJBound}, the magnitudes of the entries of the Jacobian matrix of $\etav(\rv)$ with respect to $\rv$ are uniformly bounded. 
Thus, the denoiser is Lipschitz continuous.
\end{IEEEproof}

% With Theorem~\ref{theorem:denoiser_lipschitz_continuous} established, the requirements for Condition~\ref{condition:AsymptoticCharacterization} are satisfied and the analysis provided in this article is valid. 

\end{document}